\documentstyle[aps,prd,preprint,floats]{revtex}

\newbox\rotbox
\newcommand{\be}{\begin{eqnarray}}
\newcommand{\ee}{\end{eqnarray}}
\newcommand\tag{\hbox to hsize}

\newcommand\ie {{\it i.e. }}
\newcommand\eg {{\it e.g. }}

\newcommand\del {\partial}
\def\slash#1{\rlap/{#1}}

\begin{document}
\draft

\noindent Submitted to {\it Nuclear Physics A}  \hfill  DOE/ER/40762-017

\noindent{\it }\hfill U. of MD PP\#94--037

\title{Chiral Quark Dynamics in Dense Nuclear Matter$^\dagger$}

\author{Hilmar Forkel}

\address{Department of Physics, University of Maryland, }
\address{College Park, Maryland 20742-4111 (U.S.A.)}

\date{ March 1994}
\maketitle

\begin{abstract}

We consider a new approach to the description of dense nuclear matter
in the framework of chirally symmetric, quark-based hadron models. As
previously in the Skyrme model, the dense environment is described in
terms of hyperspherical cells of unit baryon number. The intrinsic
curvature of these cells generates a new gauge interaction for the
quark fields which mediates interactions with the ambient matter.
We apply this approach to the Nambu-Jona-Lasinio (NJL)
model, construct its curved-space quark propagator and solve the ladder
Bethe-Salpeter equation for the pion. We find a high-density phase
transition to chiral restoration, discuss the density dependence of
the chiral order parameter and of the pion properties, and compare
with results of the conventional chemical-potential approach. The
new approach can additionally describe baryon-density-free cavities
in the dense medium.

\end{abstract}

\pacs{}

\narrowtext

\section{Introduction}

Nuclear matter has been the subject of intense study over the last
decades \cite{nucmat}. It represents in many ways the simplest nuclear
many-body system and also plays significant roles in other areas of
physics, ranging from relativistic heavy-ion collisions to the structure
of dense stellar objects and to the evolution of the universe. In
recent years, nuclear matter under extreme conditions, \ie at densities
a couple of times beyond the saturation density and at finite temperature,
became a particularly active research area.

Under these extreme conditions, the strong interactions and the hadron
spectrum are expected to undergo qualitative changes and, in particular,
transitions to new phases and vacua \cite{lee}. Lattice simulations
\cite{lat} and different models \cite{ber87,Kun85} predict,
for example, chiral symmetry restoration and deconfinement
transitions. The new phases will probably still be complex and
nonperturbative, with an excitation spectrum containing colorless
bound states carrying hadron quantum numbers \cite{det84} and other
collective modes, e.g. plasmons and plasminos \cite{pis}. Pion
\cite{picon} and kaon \cite{kap86} condensation and a vector-symmetric
vacuum \cite{geo90,for91} have also been contemplated. Produced in
finite volumes, some of the new vacua might furthermore have
interesting coherent decay modes \cite{nel87,bjo93}. The study of all
these phenomena promises new and direct insight into the elusive
nonperturbative sector of the underlying theory, QCD.

Since the advent of relativistic heavy-ion accelerators at Brookhaven
and CERN in 1986, matter under extreme conditions became also
accessible in the laboratory. In central heavy-ion
collisions very dense nuclear matter, albeit probably not close to
equilibrium, can be produced. Fixed-target experiments with
heavy beams, as \eg the now operational gold beam at the Brookhaven
AGS and the planned lead beam at CERN's SPS, will produce
particularly large final state multiplicities of a few thousand
charged particles per central collision, and accordingly very large
baryo-chemical potentials.

Realistic calculations of nuclear matter properties at these high
densities are difficult to perform. Direct information from the
lattice cannot be expected soon, since a chemical potential renders
the euclidean action of QCD complex and conventional Monte-Carlo
techniques inapplicable \cite{bar90}. The traditional approaches,
as for example Brueckner--Hartree--Fock and variational calculations,
are based on effective interactions (\eg from meson--exchange models
\cite{bonn}) between point-like nucleons. At high densities, however,
this neglect of the nucleon structure is a serious shortcoming, since
the chemical potential in compressed nuclear matter is of the order of
or larger than the lowest nucleon excitation energies.

Additionally, density-induced modifications of the hadron structure,
which recently became an active field of study \cite{bro87}, should be
taken into account and will in turn affect the matter properties.
Particularly dramatic changes in the hadron structure can be expected in
the vicinity of phase transitions, where the ground state of the matter
rearranges itself in a fundamental way. The chiral phase transition alters
{\it e.g.} the hadron spectrum drastically, and the deconfinement
transition transforms hadrons either into weakly bound states
\cite{det84} or dissolves them completely into a quark-gluon plasma.

To account for the hadron structure and its modifications in
high-density nuclear matter calculations is a challenging task.
Indeed, a realistic treatment is beyond the reach of present
capabilities. One can, however, attempt a more qualitative
description in the framework of baryon models, based on the
unit-cell method. This type of mean-field approach is familiar
in solid state physics for the description of homogeneous or
periodic media and becomes reliable at very high densities.
Also, it is simple enough to be applicable in many existing
baryon models.

Let us briefly recapitulate the basic concepts underlying the unit
cell approach, which we will later generalize. After dividing the
nuclear matter into identical cells of unit baryon number, periodic
(or ``twisted'' periodic) boundary conditions are imposed on the cell,
\ie the physics on opposite surfaces is identified.  As a consequence
the cell looses its boundary and becomes self-contained, while the
presence of the neighbors is reflected in its periodic images. In
topological terms, the cell becomes a three-torus $S^1 \times S^1
\times S^1$.

The size and shape of the unit cell is found by minimizing
its energy under variations of its geometry. In this way interactions
with its surroundings determine via the variational principle the
detailed cell structure. This is exactly how one derives, {\it e.g.},
the simple cubic unit cell of sodium chloride from the Coulomb
interaction. Information about the neighbors and the mutual
interactions is then encoded in one single, isolated cell. In the
context of some baryon models, and soliton models in particular,
the unit cell approach has the additional advantage of describing the
density-dependent structure of the baryons and their mutual
interactions in matter consistently, namely on the basis of the same
underlying dynamics.

In the Skyrme model \cite{BrZ87} -- a widely studied soliton model
based on an effective chiral meson lagrangian -- a face-centered
cubic (fcc) unit cell was found along these lines a few years ago
\cite{Jac88}. More recently, Manton and Ruback pointed out that the
cell energy could be lowered further by giving the unit cell an
intrinsic curvature \cite{MRu87,man87}. Their proposal can be
regarded as an extension of the variational principle, which now
allows {\it all} the geometrical properties, not only the shape, of
the cell to be determined by the given dynamics. The optimal cell
turned out to be a three-dimensional hypersphere $S^3(L)$ of radius
$L$. Exclusively in this geometry the skyrmion can attain its
absolute energy minimum, the Bogomolnyi bound \cite{MRu87}.

The simple form of the optimal unit cell, $S^3$, is essentially
determined by chiral symmetry. $S^3$ is invariant under the group
$SO(4)$, which is (locally) equivalent to the chiral group $SU(2)
\times SU(2)$. The pion fields therefore take values on $S^3$, and
their lowest-energy configuration is attained if they are defined
on another $S^3$, since the resulting map minimizes the deformation
energy stored in the field gradients.

The most remarkable new feature of hyperspherical cells is their
curvature, which induces additional dynamics for the fields. According
to the unit cell concept, these interactions are ascribed to the dense
environment. As in conventional unit cells, the cell boundary has
disappeared. The periodicity in the three angular coordinates of $S^3$
can be regarded as a reflection of the neighboring, identical cells.
The replacement of the surrounding cells and their boundaries by
periodic coordinates is, as pointed out before, a generic feature of
most unit cells and it supports the physical intuition in dealing with
the hypersphere.

The hypersphere approach has by now been extensively applied in the
Skyrme model. Perhaps most remarkably, a chiral phase transition
emerged at high densities, accompanied by parity doubling of the
hadron spectrum and the disappearance of the Goldstone modes
\cite{for289}. Also, strangeness condensates with distinct
characteristics
due to changes in the extended baryon structure have been found
\cite{for189}, and the vector limit \cite{geo90}  phase transition
has been studied \cite{for91}. Furthermore, bag formation and
dissolution \cite{Rei87}, some properties of the loop expansion in
the presence of a skyrmion \cite{Wil88} and its excitation spectrum
\cite{wir90} have been investigated. Skyrmion-like profiles on the
hypersphere have also been derived from instantons \cite{Tem88},
following an idea by Atiyah and Manton \cite{ati89}, and other
soliton solutions have been constructed \cite{jac89}.

A remarkable and generic finding of these studies is that hypersphere
calculations reproduce almost quantitatively the results of similar
fcc array calculations, whenever those are available. This holds in
particular for the chiral phase transition, its critical density,
the density dependence of the order parameter and the energy
density, and it established confidence in the new approach. At the
same time the calculational effort is greatly reduced. Array
calculations require the numerical solution of partial-differential
field equations on a three-dimensional grid. Hypersphere calculations,
on the other hand, involve at most the solution of an ordinary
differential equation, as for isolated skyrmions. Some results can
even be obtained analytically and allow a straightforward and
transparent physical interpretation.

The rather successful and consistent results of this unorthodox
approach call for both a better understanding of its physical
foundations and for further exploration of its range of
applicability. In the present paper we will take a first step
in this direction by extending the hypersphere approach to
quark-based hadron models, and to the Nambu-Jona-Lasinio (NJL)
model \cite{NJL69} in particular. Some results in the chiral
symmetry breaking sector have already been published in ref.
\cite{for92}.

The NJL model shares with the Skyrme model the underlying chiral
symmetry, which led to the geometry of $S^3$. In many other respects,
however, the NJL dynamics is different. In contrast to the purely
mesonic degrees of freedom in Skyrme-like models, it is based solely
on quark fields. Furthermore, it does not describe nucleons as
solitons and it breaks chiral symmetry dynamically. These
differences can help to disentangle effects specific to the
Skyrme model from more generic or even model-independent features
of the hypersphere approach. In particular, they could help to
clarify whether the specific geometrical and topological properties
of the Skyrme model are indispensable for the effectiveness of the
hypersphere approach.

The choice of $S^3$ unit cells even beyond the Skyrme model is
additionally motivated by their maximal symmetry
under all three-dimensional curved spaces. Hyperspherical cells
can thus be considered as the simplest generalization of flat
unit cells, and many of their characteristic features are indeed
reminiscent of flat cells. This tendency to preserve qualitative
properties of euclidean space is enhanced by the conformal
equivalence between $S^3$ and $R^3$. The use of $S^3$ cells
furthermore avoids potential problems with explicit chiral
symmetry breaking by a cell boundary (as, for example, in bag
models).

An alternative, more conventional approach to dense matter in
the NJL model, in which the quarks are coupled to a homogeneous,
external baryon number source, has been studied by Bernard et
al. \cite{ber87} and by others \cite{NJLdens}. As in the
Skyrme model, where the comparison with conventional array
calculations served as a test for the hypersphere approach, we
will profit from comparing our results to those of the chemical
potential approach.

Let us add a couple of comments concerning the uses and
limitations of our approach. Any attempt to describe nuclear
matter in terms of unit cells, \ie in a mean field framework,
is bound to be qualitative at best \cite{coh89}. Approximating
the short- and medium-range nucleon-nucleon correlations by a
strict long-range order and neglecting the nucleons kinetic
energy are of course rather severe simplifications, except at
very high densities such as in neutron stars, where nuclear
matter might even crystallize.

On the other hand, one may expect that bulk features of nuclear
matter are robust enough to survive these approximations, and
that they can be revealed in the unit-cell framework. Global
features and driving mechanisms of high-density phase transitions,
for example, could fall into this category, and the alternative
Fermi-gas treatment will fail to describe them. As already
mentioned, the unit cell approach offers the so far unique
possibility to account consistently for the extended baryon
structure, which certainly contributes qualitatively important
physics at high densities.

The paper is organized as follows: In section II we establish the
general framework for our investigation by implementing the quark
fields into the curved unit cell. Section III deals with the
specific NJL dynamics and the construction of its curved-cell
quark propagator in Hartree-Fock approximation. The latter is a
key step in our program. Its more technical parts are relegated
to appendices A and B.

In section IV the chiral symmetry breaking sector and its density
dependence is examined. We derive, in particular, the gap equation,
which describes the dynamical quark mass generation, and the
expression for the chiral order parameter. In section V the
Bethe-Salpeter equation in the quark-antiquark channel is set up
in the curved background, and its solution for the pion in ladder
approximation is found.

{}From the explicit pion wave function we calculate in section VI
the pion decay constant and its density dependence. In section
VII the zero-density limits of the various calculated quantities
are derived and compared with the standard flat-space NJL results.
Section VIII discusses the quantitative aspects of our results and
compares them with results of the chemical potential approach.
Finally, section IX summarizes the paper and offers some
conclusions and perspectives for future research.

\section{Quarks in the curved unit cell}

In this section we describe the implementation of quark fields
into a curved unit cell and summarize the generic physical
implications. This discussion is quite general and provides
a framework for the adaptation of a large class of quark-based
baryon models.

As motivated in the introduction, our unit cell has the geometry
of a three-dimensional hypersphere, $S^3(L)$, of radius $L$.
Accordingly, the flat Minkowski metric is replaced by the metric
of $S^3(L) \times R$ ($R$ indicates the real, flat time
dimension),
\be
ds^2 = g_{\mu \nu} dx^{\mu} dx^{\nu} = dt^2 - L^2 [d\mu^2 +
\sin^2\mu
(d\theta^2 + \sin^2\theta d\phi^2)], \label{metric}
\ee
which we write in the polar coordinates of $R^4$,
\ie $\{ x^{\mu} \}= (t,\mu,\theta,\phi)$. Its intrinsic curvature
leads to new interactions for the spinor fields, as it does, for
example, in general relativity. In our context, however, these
interactions will be ascribed to the dense environment.

The spatial section of (\ref{metric}) has six symmetry
generators\footnote{These $SO(4)$ generators form
together with the time translation generator the Killing vectors
of the metric (\ref{metric}).}, the maximal number in a
three-dimensional, curved space. Three of them are just the ordinary
rotations in $R^3$, while the remaining three are compact analogs
of Lorentz boost generators and can be regarded as ``generalized
translations'' in $S^3$. Combined with the usual translations in
the flat time direction, the metric (\ref{metric}) therefore
possesses seven symmetries, only three less than the Minkowski
metric of $R^4$. This large symmetry group will be helpful both
in establishing analogies with flat unit cells and in explicit
calculations. Together with the conformal relation between the
hypersphere and Minkowski space, it often leads to structural
similarities in expressions for physical quantities.

In the Skyrme model, which contains only spin-0 pion fields, the
change of the lagrangian to the new metric was sufficient
to adapt the model to the hypersphere. Spinor fields in curved
space, however, require the introduction of additional concepts.
The straightforward generalization of the fields from representations
of the Lorentz group to representations of the general linear group
of (infinitesimal) coordinate transformations fails for fermions,
since the linear group has no spinor representations.

This impasse can be circumvented \cite{Egu77} by defining the
spinor fields in a {\it local} orthonormal basis, given by the
vierbein fields $e^a(x)$. The $e^a$ are generated by linear,
space-time dependent transformations of the coordinate basis,
written in terms of the $4 \times 4$ matrix $e^a_{\,\,\,\mu} (x)$:
\be
e^a(x) = e^a_{\,\,\,\mu} (x) \,dx^{\mu}, \qquad \qquad a \in
\{ 0,1,2,3 \}. \label{vbein}
\ee
The corresponding vector fields, which form the components of the
gradient operator in the orthonormal basis, can be obtained from the
inverse vierbein coefficients $e_b^{\,\,\,\mu}$, defined by
\be
e^a_{\,\,\,\mu} (x) \, e_b^{\,\,\,\mu} (x) = \delta^a_b.
\ee
The $e_b^{\,\,\,\mu}$ transform coordinate-vector components into
Lorentz-vector components (in the local frame) and are obtained from
the $e^a_{\,\,\,\mu}$ by lowering the Latin index with the Minkowski
metric  $\{ \eta_{ab} \} = diag(1,-1,-1,-1)$ of the local frames
and by raising the Greek index with the hypersphere metric $g$. The
gradients are then given by
\be
e_a(x) = e_a^{\,\,\,\mu} (x) \,\, \del_{\mu}. \label{vbein2}
\ee
Due to its orthonormality, the vierbein is simply related
to the metric (\ref{metric}):
\be
g_{\mu\nu} (x) = \eta_{ab} \,\, e^a_{\,\,\,\mu} (x) \,
e^b_{\,\,\,\nu} (x). \label{metviel}
\ee
In our context it is crucial to note that this decomposition of a
given metric in terms of the vierbein frames is not unique. Indeed,
an arbitrary, {\it local} Lorentz transformation of the Latin vierbein
indices leaves the expression (\ref{metviel}) unchanged and thus
leads to the same metric. These Lorentz rotations consequently form a
gauge group, which just changes the relative orientation convention
(the ``gauge'') of the local frames, but leaves the physics (i.e. the
metric) invariant. It generalizes the global Lorentz group of Minkowski
space, and its spinor representations allow the definition of fermion
fields in curved space. In order to preserve the local Lorentz invariance
of the lagrangian, gradients of the quark fields have to be
replaced by gauge-covariant derivatives,
\be
e_a^{\,\,\,\mu} (\partial_{\mu} +  \omega_{\mu}) \, q \equiv
( e_a +  \omega_a ) \, q , \label{codev}
\ee
which transform homogeneously under the gauge group. In eq. (\ref{codev})
we have introduced a ($4 \times 4$-matrix-valued) gauge field, the spin
connection
\be
\omega_a =  e_a^{\,\,\,\mu} \,\, \omega_{\mu} \equiv -\frac{i}{4}
\omega^{\,\,\,bc}_a \sigma_{bc}
\ee
($\sigma_{ab} = \frac{i}{2} [\gamma_a,\gamma_b]$), which just
compensates the change of the spinor field components due to different
frame orientations at neighboring points by a spatial $SO(3)$ rotation.
In this paper, we take $\omega$ to be the Levi-Civita
connection\footnote{The Levi-Civita connection is metric and
torsion-free (see appendix A) and thus corresponds to the natural
and minimal generalization of flat unit cells. In principle, one
could use a more general spin connection, which would introduce a
non-zero torsion into the unit cell. This issue might deserve
future investigation. }, which is uniquely determined by the
metric and the vierbein  \cite{Egu77}:
\be \omega^{\,\,\,bc}_a  = - e^b_{\,\,\,\nu} e^c_{\,\,\,\rho}
\left( \delta^{\rho}_{\mu} \, \del^{\nu} - \Gamma^{\,\,\,\nu \rho}_{\mu}
\right) e_a^{\,\,\,\mu}.
\ee
The Christoffel connection $\Gamma$ is the corresponding gauge field of
the local Lorentz group in the vector representation and can be directly
obtained from the metric:
\be
\Gamma^{\,\,\,\nu \rho}_{\mu} = \frac{1}{2} g_{\sigma \mu}
(\del^{\nu} g^{\rho \sigma}  + \del^{\rho} g^{\nu \sigma} -
\del^{\sigma} g^{\nu \rho}). \label{crist}
\ee

The explicit form of $\omega$ depends, of course, on the gauge. The
group structure of $S^3$ implies the existence of a particularly
useful choice, which we will refer to as the ``Maurer-Cartan gauge''.
In it, the spin connection takes a simple and, in particular,
space-time independent form:
\be
\omega_a = \frac{i}{4L} \epsilon_{abc} \sigma^{bc} \quad (a,b,c
\,\,\epsilon \,
\{ 1,2,3 \} ) \qquad {\rm and} \quad \omega_o = 0.
\ee
($\sigma_{ab} = \frac{i}{2} [\gamma_a,\gamma_b]$) More details about
this gauge, in which all of the following calculations will be
performed, can be found in appendix A.

In the context of nuclear matter unit-cells, it might at first be
tempting to think
of the spin connection as a covariantized chemical potential, since it
acts as a constant, isoscalar vector interaction for the quarks and
grows with density. Such an interpretation would, however, be misleading.
Besides its gauge dependence, $\omega$ has a different Dirac-matrix
structure and its time component vanishes identically in the matter
rest frame.

Following the procedure outlined above, one can now adopt a general
fermionic lagrangian from Minkowski space to the hypersphere. The
corresponding action is obtained by integrating the lagrangian
over $S^3(L) \times R$, \ie
\be
S = \int d \mu (x) \,\,  {\cal L} (x). \label{action}
\ee
The integration measure $d \mu(x)$ contains the jacobian
$\sqrt{ - {\rm g}}$, where ${\rm g}(x) = -L^6 \sin^4 \mu \sin^2
\theta$ is the determinant of the metric tensor (\ref{metric}):
\be
d \mu(x) =  \sqrt{ - {\rm g} (x) } \, d^4 x  =   L^3 \sin^2 \mu
\sin \theta \, d\mu \, d \theta \, d \phi \, d t .  \label{measure}
\ee

In order to complete the definition of the unit cell it remains to
set its total baryon number to one. Since this is most conveniently
done at the level of the quark propagator, we will reserve this last
step for the next section.

Let us now summarize the model-independent physical implications
of the curved unit cell for fermions. The novel aspects can be
classified according to their either local or global nature.
The local effects are due to the finite curvature: gradients are
replaced by vierbein vector fields and a new gauge interaction
emerges.

However, as already pointed out in the
introduction, there are also new features of global, \ie topological
character. Due to the compactness of the hypersphere, the fermion
spectrum (see appendix A for its explicit form) will be discrete,
with the energy scale of the excitations determined by the inverse
radius of the cell. Furthermore, potentially symmetry-breaking cell
boundaries are absent. The periodicity in the three (angular) spatial
directions $\mu, \theta$ and $\phi$ is reminiscent of the periodic
boundary conditions in flat unit cells and can be regarded as a
reflection of the neighboring, identical cells. Quarks
moving around the cell more than once are accordingly interpreted
as entering a neighboring cell. This physical interpretation will
become explicit in particular during the construction of the
fermion propagator.

We conclude this section with a remark on the perturbative
treatment of space curvature, which is frequently used in general
relativity. While this approach often significantly simplifies
explicit calculations, it will be insufficient for our purpose,
for two reasons: First, the curvature of our unit cell is of the
order of $V^{-\frac{1}{3}}$, where $V$ is the average volume
occupied by one nucleon in nuclear matter, and thus becomes large
at high densities. Secondly, a perturbative approximation to the
metric disregards the global structure of the unit cell.
This is another serious  shortcoming at high densities, where the
cell becomes smaller and, as we will show in the next section,
the quarks get lighter on the approach to chiral restoration.
Under these circumstances the global features of the cell become
important.

\section{NJL model and quark propagator on $S^3 \times R$}

Up to now our discussion established a generic framework for the
study of fermionic models in hyperspherical unit cells. In order
to proceed further and to arrive at quantitative results, we
have to specify the dynamics of the quarks.

Many of the existing, chirally symmetric quark and quark-meson
models are accessible in our approach. The Nambu-Jona-Lasinio
(NJL) model \cite{NJL69}, however, seems particularly well suited
for a first, explorative study aimed at testing the use of
hyperspherical unit cells beyond the Skyrme model. As already
indicated in the introduction, the NJL dynamics is simple and
contrasts the Skyrme model in many aspects, except for their
common chiral symmetry. For example, the NJL model is entirely
quark-based (in its modern form), it is not a soliton
model\footnote{In this paper we will consider only the original,
fermionic version of the NJL model. A restricted  bosonization of
some generalized NJL lagrangians can support stable, skyrmion-like
soliton solutions \cite{bos}. Their eventual investigation on $S^3$
could shed more light on connections between the hypersphere
results in the Skyrme and NJL models. } and it breaks chiral symmetry
dynamically.

In flat Minkowski space, the flavor-$SU(2)$ version of the NJL
model \cite{NJL69} is based on the lagrangian
\be
{\cal L} = \frac{i}{2} \left[ \, \overline{q} \gamma^{\mu}
\partial_{\mu} q -
(\partial_{\mu} \overline{q}) \gamma^{\mu}  q \, \right]  -
m_o \overline{q}
q +  g \hspace{1mm} (\overline{q} \Gamma_a  q)  \hspace{1mm}
( \overline{q} \Gamma_a q). \label{NJL}
\ee
The $q$'s denote Dirac-spinor quark fields, which are color triplets
and isospin doublets. The scalar and pseudoscalar four-quark contact
interactions resemble instanton-generated vertices of light
quarks in QCD \cite{tHoo76} and are written in a shorthand notation
with the help of the four matrices $\Gamma^o = 1_{\gamma} 1_{\tau},
\,\,\, \Gamma^i =  i \gamma_5 \tau_i \,\,\,(i=1,2,3)$. (The index $a
\in \{0,1,2,3\}$ is implicitly summed over in the interaction term
above.) In most of the following discussion we will set the small
current quark masses $m_o$ to zero.

Due to its dynamical symmetry breaking mechanism, simplicity and
phenomenological successes, the NJL model is widely employed as an
effective low-energy theory for QCD \cite{vog91,klev92}. Recently,
the original version of the NJL model has been generalized
in a couple of directions, \eg by adding vector and axial-vector
couplings or by extensions to the $SU(3)$-flavor sector. These
developments and many applications to meson and baryon physics,
including some in hot and dense nuclear matter, are reviewed in
refs. \cite{vog91,klev92}.  We will work with the original,
``minimal'' version of the model, however, since our aim is not
a detailed phenomenological analysis, but rather the investigation
of generic aspects of the new approach.

We begin our study by adapting the NJL lagrangian (\ref{NJL})
according to the procedure established in the preceding section
to the hyperspherical unit cell, \ie we replace the flat
Minkowski metric by eq. (\ref{metric}) and the gradients by
gauge-covariant derivatives:
\be
{\cal L} = \frac{i}{2} \left[\, \overline{q} \gamma^a ( e_a +
\omega_a) q -
\overline{q}( \overleftarrow{e}_a - \omega_a) \gamma^a q \, \right]
\,\, - m_o
\overline{q} q + g \, (\overline{q} \, \Gamma_a  q)
\, (\overline{q} \, \Gamma^a q). \label{njlhyp}
\ee
(The arrow indicates that the corresponding differential operator
acts to the left.)

Most of our following analysis will be based on the standard
Hartree-Fock approximation to the NJL dynamics \cite{NJL69}.
As is well known, the quarks dynamically acquire already at
this level a finite self-energy, if the coupling $g$ exceeds a
critical value. Due to the pointlike character of the interaction,
the self-energy is space-time independent. The additional gauge
interaction on the hypersphere preserves this property, since the
spin connection is a non-propagating background field. In the
presence of a finite baryon density, the self-energy acquires
besides the standard scalar piece, which acts as a  ``constituent''
mass, an additional vector part:
\be
\Sigma = \Sigma_s - \gamma_0 \Sigma_v . \label{selfen1}
\ee

All the vacuum, quark and pion properties of interest in this
paper can be conveniently calculated with the help of the NJL
constituent quark propagator, which includes the self-energy
(\ref{selfen1}). The next and pivotal step of our program is
therefore the derivation of an explicit expression for this
propagator {\it on $S^3 \times R$}.

To this end, and at the densities of interest the crucial
modifications in the quark dynamics due to the large curvature
have to be taken into account exactly. While in general
analytical expressions for
propagators in curved spaces can rarely be found, the high symmetry
of the hypersphere allows us to derive the quark propagator on
$S^3\times R$ exactly and in closed form. Our derivation starts from
the definition of the propagator as the inverse of the gauge-covariant
Dirac operator,
\be
\left[  i \gamma^a ( e_a + \omega_a) + \Sigma_v \gamma_0 -
(m_0 + \Sigma_s) \right] \, S(x,y)  = \delta^4 (x,y).
\ee
(The explicit form of the delta function on $S^3 \times R$ is given
in appendix B.) We then specialize to the Maurer-Cartan gauge,
separate $S$ into invariant scalar amplitudes and rewrite them such
that the dependence on all but the geodesic coordinate of $S^3$
is removed. The remaining $\mu$ dependence can then be absorbed
into redefined amplitudes, which finally allows the inversion to be
performed by Fourier methods. The detailed calculation is described
in appendix B.

The resulting Hartree-Fock propagator on $S^3 \times R$, in the
chiral $\gamma$-matrix basis, reads
\be
S(x) \equiv S(x,0) = {\rm e}^{ i ( \vec{ \Sigma } \hat{r}
\frac{\mu}{2} + \Sigma_v t)} \, [ \,S_0 (\mu,t) \, \gamma_0 - S_1
(\mu,t) \, \hat{r} \vec{\gamma} - S_2 (\mu,t) \, ]. \label{fullprop}
\ee
The exponential factor, which contains the Dirac spin
matrix\footnote{$\vec{ \Sigma }$ is the direct product of the
$2 \times 2$ unit matrix and the standard Pauli matrices, see
appendix B.} $\vec{ \Sigma }$, is the spin parallel propagator
and $\hat{r} (\theta, \phi)$ denotes the usual unit-vector of
euclidean $R^3$ in spherical coordinates. The decomposition of
the propagator into the three amplitudes
\be
S_0 (\mu,t) &=& -  i \alpha \, (\sin \mu)^{-1} \,
\sum_{n=-\infty}^{+ \infty} (-1)^n [2 \dot{I}_n' + \tan
\frac{\mu}{2} \dot{I}_n ], \label{SS0} \\
S_1 (\mu,t) &=&  i \, \frac{\alpha }{L} \, (\sin \mu)^{-1}
\, \sum_{n=-\infty}^{+ \infty} (-1)^n [2 I''_n - \cot \frac{\mu}{2}
I'_n], \label{SS1} \\
S_2(\mu,t) &=& m \alpha \, (\sin \mu)^{-1} \,
\sum_{n=-\infty}^{+ \infty} (-1)^n [2 I_n' + \tan \frac{\mu}{2} I_n ]
, \label{SS2}
\ee
($\alpha \equiv 1/ 4 \pi L^2$) reflects the symmetry properties of
$S^3 \times R$. The $S_i (\mu,t)$  are expressed as sums over all
geodesic paths on which a quark can propagate between the pole
($y=0$) and the point $x$. Since the cell is compact, these paths
contain $n$ full circles around the hypersphere. This becomes
transparent by noting that all the integrals $I_n \equiv I (\mu +
2n \pi,t)$ and their derivatives with
respect to $\mu$ ($t$), denoted by primes (dots),
are derived from the two-dimensional propagator
\be
I(\mu,t) = \int \frac{dk_0}{2  \pi} \int \frac{d k}{2\pi} \,\,
\frac{{\rm e}^{i(k\mu L - k_0 t)} }{k_0^2 - k^2 - m^2 + i\epsilon}
\label{2dprop}
\ee
($m =  m_0 + \Sigma_s$). The additional $\mu$ dependence in equations
(\ref{SS0}) -- (\ref{SS2}) just translates propagation along a flat
``$\mu$-axis'' into geodesic propagation in the $\mu$ direction on
the curved hypersphere.

The propagator (\ref{fullprop}) describes quark propagation in a
cell with zero net baryon number and consequently satisfies the
usual Feynman vacuum boundary conditions. Its adaptation to the
cell's physical ground state with baryon number one
is, however,  straightforward, owing to the transparent
implementation of the boundary conditions in eq.
(\ref{2dprop}).

The ``Fermi sea'' of three valence quarks in the cell affects
the singularities of the integrand of $I(\mu,t)$ inside the Fermi
sphere both by the appearance of poles from positive energy holes
and by the suppression of poles from Pauli-blocked states.
Accordingly, the integrand in eq. ({\ref{2dprop}) has to be modified
to
\be
e^{i(k\mu L - k_0 t)} \left[ \frac{1}{k_0^2 - \omega^2 + i\epsilon}
+ \frac{ i \pi}{\omega} \delta (k_0 - \omega) \Theta (k_F -
|k|) \right] ,
\ee
where $\omega = \sqrt{k^2 + m^2}$ and $k_F$ is the Fermi momentum
of the quarks. With the explicit expression for the propagator at
hand, we are now ready to enter the discussion of the quark dynamics
in the hyperspherical cell.

\section{Dynamical chiral symmetry breaking}

In the Skyrme model, the perhaps most important and unexpected result
of the hypersphere approach was the prediction of a chiral
restoration phase transition at high densities \cite{man87,for289}.
However, the restoration mechanism exploits specific topological
and geometrical features and, in particular, the hedgehog structure
of the skyrmion\footnote{In the Skyrme model in flat space, chiral
symmetry is broken by a nonlinear parametrization of the pion fields.
On the hypersphere at high densities the ground state
changes into a generalized hedgehog form. In addition to the coupling
of isospin transformations and rotations of the standard hedgehog, it
also couples the axial generators of the chiral group to the
``translations''
on the hypersphere. Projection onto the physical rotation {\it and}
translation eigenstates then reveals the restoration of the full
chiral symmetry \cite{for289}.}. It is therefore not {\it a priori}
clear if and how hyperspherical unit cells can give rise to chiral
restoration in other models.

In the NJL model, in particular, chiral symmetry breaking occurs
dynamically, by quark-antiquark pair condensation in the vacuum
\cite{NJL69}. At the same time the quarks become ``dressed'' by
a virtual quark-antiquark cloud and acquire a ``constituent'' mass.
This mechanism, analogous to the BCS mechanism for gap formation in
superconductivity and perhaps mediated in a similar way by instantons
in QCD \cite{ShuVer90}, is fundamentally different from its Skyrme
model counterpart. In the present section we will examine how it
is affected by a finite baryon density, described in terms of
hyperspherical unit cells.

We base our study, following Nambu and Jona-Lasinio, on the
Schwinger-Dyson equation
\be
\Sigma (x,y) = S_0^{-1} (x,y) - S^{-1} (x,y) \label{schwi}
\ee
for the quark self-energy, since the development of a constituent
quark mass is a direct signature for chiral symmetry breaking.
Equation (\ref{schwi}) has the same form as in flat space and its
derivation proceeds along the same standard lines \cite{Itz80}.
Note, however, that both $S$ and its free (\ie $g = 0$)
counterpart $S_0$
contain the interactions with the curved background and, in
particular, with the gauge field $\omega$.

It is well known that nontrivial solutions of eq. (\ref{schwi}),
{\it i.e.} finite constituent masses, can only be generated
nonperturbatively. Both the Hartree approximation, which becomes
exact in the large-$n_c$ limit of the NJL model ($n_c$ is
the number of colors of the quark fields), and the Hartree-Fock
approximation are frequently used for this purpose. In the following,
we will adopt the Hartree-Fock approximation to allow for a direct
comparison with results of the conventional approach \cite{ber87}.
To find the Hartree-Fock solution of eq. (\ref{schwi}), we start
from the solution to $O(g)$, \ie to first order in the NJL
interaction:
\be
\Sigma(x,y) =  \Sigma (x) \, \delta^4(x,y),
\ee
with
\be
\Sigma (x) \equiv \Sigma =  i g \,  \{  \, 2 \Gamma_a \,
{\rm tr} \, [S_0^- (x,x) \Gamma_a ]  - \Gamma_a \, [S_0^+ (x,x)  +
S_0^- (x,x) ] \, \Gamma_a \,  \} . \label{sol}
\ee
Due to the pointlike character of the NJL interaction the self-energy
is, as anticipated, space-time independent. For the same reason, both
coincidence limits of the propagator (\ref{fullprop}) appear above,
\be
S_0^{\pm} (x,x) = S^{\pm} (0) =   \lim_{x_0 \rightarrow 0 \pm}
S(x_0, {\bf x} = 0) = \tilde{S}_0^{\pm} \gamma_0 - \tilde{S}_2,
\label{coin}
\ee
and are independent of $x$. Note that the amplitude $S_1$ in eq.
(\ref{fullprop}) has to vanish in the coincidence limit, since it
multiplies the unit distance vector.

Both $S_0$ and $S_2$ contain (quadratic and logarithmic) short
distances singularities, which originate from the integral $I(\mu,t)$,
eq. (\ref{2dprop}), and its derivatives. We therefore multiply the
corresponding integrands by a smoothed theta
function\footnote{Due to its negative mass dimension, the NJL
coupling is not renormalizable. The introduced cutoff $\Lambda$
is therefore physical \cite{Ber88}.}
\be
\Theta_{\epsilon} (\Lambda - |k|) = \frac{1}{{\rm e}^{\epsilon L
(|k| - \Lambda)} + 1}, \label{cutoff}
\ee
which damps the contributions from the high-momentum region and
thus regularizes the integrals. Note that cutting off the spatial
part
of the momentum integration is consistent with the symmetry
properties of the metric and of the ``Fermi sea'' of valence
quarks.

The representation of the propagator amplitudes as a sum over paths,
eqs. (\ref{SS0}) --  (\ref{SS2}), is sometimes inconvenient for
practical purposes. At high densities, in particular, the unit
cell radius $L$ is small and a large number of paths contribute.
In appendix C, these sums are therefore transformed into equivalent
mode sums over the discrete Dirac spectrum on $S^3 \times R$, with
the energies
\be
\omega_n = \sqrt{k_n^2 + m^2}, \quad \quad \qquad k_n =
\frac{ 2n + 1}{2 L},
\quad \qquad n \in \{1, 2, 3, ...\}
\ee
and the corresponding degeneracies
\be
D_n = 2n (n+1) \label{degen1}
\ee
of the $n$th quark level (for a given flavor and color), which
emerges in the course of the transformation. (An alternative,
more direct derivation of the spectrum is given in appendix A.)
In the spectral representation, the regularized coincidence limits
of the propagator functions take the form (cf. appendix C)
\be
\tilde{S}_0^{\pm} &=& \mp \frac{  i}{4V}   \sum_{n= 1 }^{\infty}
D_n \Theta_{\epsilon} (\Lambda - k_n) \Theta (k_F - k_n) ,
\label{s0coin} \\
\tilde{S}_2 &=& \frac{  i m}{4V}  \sum_{n =1}^{\infty}
\frac{D_n }{\sqrt{k_n^2 + m^2}} \Theta_{\epsilon} (\Lambda - k_n)
\Theta (k_n - k_F), \label{s2coin}
\ee
where $V = 2 \pi^2 L^3$ is the volume of the unit cell.

{}From the Dirac spectrum we can immediately read off the Fermi energy
$\omega_F$, since all three valence quarks are contained in the
lowest
energy level:
\be
\omega_F = \sqrt{k_F^2 + m^2}, \quad \quad k_F = k_1 = \frac{3}{2L}.
\ee
Since this level is only partially filled, we have to prevent its
remaining states from being counted in the above mode sums. To this
end we associate a filling factor
\be
f_n = \frac{N_n }{n_c n_f D_n}
\ee
with each level, where $N_n$ is the number of {\it valence} quarks
in the nth level, {\it i.e.} $f_1 = \frac{1}{8}$ and $f_n = 0$
for $ n >1$.

We are now ready to give explicit expressions for the self-energy,
which properly take the valence quarks into account. Decomposing
the Dirac-matrix structure of the self-energy in the rest frame
into scalar and vector parts,
\be
\Sigma = \Sigma_s - \gamma_0 \Sigma_v , \label{selfen}
\ee
as suggested by the symmetry properties and anticipated in the last
section, and inserting the coincidence limit of the propagator into
eq. (\ref{sol}), we obtain
\be
\Sigma_s &=& - 4  i g \, \left(1+2 n_c n_f \right) \, \tilde{S}_2
\nonumber \\ &=& \frac{m g}{V} (1+2 n_c n_f)  \sum_{n=1}^{\infty}
\frac{D_n (1 - f_n)}{\omega_n}  \Theta_{\epsilon} (\Lambda - k_n),
\label{sselffinal} \\
\Sigma_v &=&  4  i g \, ( \tilde{S}_0^{+} + \tilde{S}_0^{-} )
\nonumber \\ &=& \frac{-2g}{V} \sum_{n=1}^{\infty} D_n f_n = -
\frac{g}{V}. \label{vselffinal}
\ee
The above expressions exhibit clear analogies
to their flat-space counterparts and allow a direct physical
interpretation. The scalar self-energy receives contributions from
virtual quark loops in all accessible levels of the Dirac sea
down to the smooth cutoff $\Lambda$. The three occupied states in
the Fermi sea contribute with opposite sign. In both self-energies
the finite-density effects, mediated by the curved cell, manifest
themselves mainly through the altered quark momenta $k_n$ and
degeneracies $D_n$, which replace the plane wave spectrum of flat
space.

The vector part of the self-energy is entirely due to
the three valence quarks and vanishes in an ``empty'' cell, even
at finite ambient baryon density. This is in contrast to the
chemical-potential approach of ref. \cite{ber87},
where $\Sigma_v$ is necessarily finite at finite density.
Furthermore,
$\Sigma_v$ does not depend on the value of the Fermi energy,
but only on its density-independent relative position in the
spectrum.

We note in passing that the not explicitly $n_c$-dependent
terms in the scalar self-energy (\ref{sselffinal}) arise
from exchange
diagrams and become negligible in the large-$n_c$ limit of the
NJL model\footnote{At the physical $n_c = 3$ the exchange
contributions
are an order of magnitude smaller than the direct terms.}.
They are kept, however, in the Hartree-Fock approximation.

Since a constant self-energy just shifts the mass and energy
scales
of the quarks, the $O(g)$ expressions in eqs. (\ref{sselffinal})
and (\ref{vselffinal}) have already the functional form of the
Hartree-Fock solutions. We still have to impose the
self-consistency condition, however, which requires the mass
in the quark loop-propagator on the right-hand
side of eq. (\ref{sselffinal}) to be identical to the scalar
part of the self-energy itself\footnote{Since our focus is on
spontaneous chiral symmetry breaking, we neglect the small
explicit breaking and specialize to the chiral limit, $m_0 = 0$.
The generalization to a finite current mass is straightforward.}
and leads to the gap equation
\be
m =  \frac{m g}{V} (1+2 n_c n_f)  \sum_{1}^{\infty} \frac{D_n (1-f_n)
 }{\omega_n}\Theta_{\epsilon} (\Lambda - k_n ). \label{gapeq}
\ee

At least at low densities and for sufficiently large coupling $g$,
we expect the perturbative vacuum to become unstable, as it does
in flat space. The simultaneously generated quark mass can then be
found as the nontrivial solution of eq. (\ref{gapeq}). The additional,
trivial solution $m=0$ does of course always exist.

With the self-consistent self-energies at hand, the Hartree-Fock
propagator is now completely determined and vacuum expectation values
of bilinear quark operators can be calculated. In the remainder of
this section, we will specifically consider the vector and scalar
quark condensates. The vector condensate, or equivalently the quark
number density, provides a consistency check on our treatment
of the valence quark sector, whereas the scalar condensate plays a
central role in the discussion of dynamical chiral symmetry breaking
and its density dependence.

Let us start with the vector condensates. They are equal to the quark
number densities in the cell, which add up to
\be
<u^{\dagger} u> + <d^{\dagger} d> = \frac{3}{V},
\ee
corresponding to three valence quarks per cell. We further specialize
to isosymmetric nuclear matter, so that the quark number densities
of both up and down quarks become equal:
\be
<u^{\dagger} u> = <d^{\dagger} d> \equiv <q^{\dagger} q> =
\frac{3}{2V}. \label{qdens}
\ee
To check our treatment of the valence quark sector, we
now calculate $<q^{\dagger} q>$ for comparison directly from the
coincidence limit of the quark propagator, eq. (\ref{coin}):
\be
<q^{\dagger} q> &=& -  i {\rm tr}_{\gamma,c} \left[ \gamma_0
(S^-(0)_{\rho} - S(0)_{\rho = 0} ) \right] \nonumber \\
 &=&  -  i {\rm tr}_{\gamma,c} \left[ \gamma_0 \tilde{S}_0^{-}
\right] = \frac{n_c}{V} D_1 f_1 =  \frac{n_c}{2V}.
\ee
As expected, the two results agree and thus confirm the
correct pole structure of the propagator.

We now turn to the scalar quark condensate, which
is the standard order parameter of the chiral phase transition.
Again, it can be directly obtained from the propagator:
\be
<\bar{q} q> &=& - \lim_{t \rightarrow 0} {\rm tr}_{\gamma,c}
<Tq({\bf x} = 0,t) \bar{q}(0)> = -  i {\rm tr}_{\gamma,c}
\left[ S^-(0) \right] \nonumber \\ &=& - \frac{n_c m}{V}
\sum_{n=1}^{\infty} \frac{D_n (1-f_n)}{\omega_n} \Theta_{\epsilon}
(\Lambda - k_n). \label{qcond}
\ee
Note that  $<\bar{q} q>$ is proportional to the quark mass,
{\it i.e.} to the solution of the gap equation (\ref{gapeq}). Chiral
symmetry breaking manifests itself, as anticipated, simultaneously
in a finite quark condensate and a dynamically generated quark mass.
If a nontrivial solution of the gap equation exists and if it is
energetically favorable compared to the other solutions (see the
discussion in section VIII), a quark condensate necessarily develops.

The gap equation (\ref{gapeq}) can be solved numerically and indeed
yields, at low densities and for sufficiently large couplings, a finite
quark mass. As expected, the mass decreases with density and finally
vanishes, signaling the chiral restoration transition. The quantitative
solution of the gap equation as a function of density and the evaluation
of the associated free energy will be subject of section VIII, where
we also discuss the behavior of the scalar quark condensate and the
order of the chiral phase transition. Section VIII furthermore
contains
a comparison of our results with those of ref. \cite{ber87}.

\section{The pion wave function}

As a consequence of Goldstone's theorem, the spontaneous breakdown
of chiral symmetry is accompanied by the appearance of an iso-triplet
of (almost) massless pions. They emerge in the NJL model as
quark-antiquark pairs, tightly bound by the same interaction which
leads to quark condensation. In the present section we study this
mechanism and its density dependence in our framework.

Our strategy will be to generalize the original NJL approach of
calculating the pion wave function to the curved unit cell. Most
of the new features encountered are related to the changes in both
the quark and pion spectra. (For the actual spectra of free fermions
and spin-0 bosons on $S^3 \times R$ see appendices A and D.) As
a consequence, the Fourier transform of the spatial coordinate
dependence becomes impractical and our calculation will be done
in coordinate space.

Following NJL \cite{NJL69}, we start from the connected four-point
function in the quark-antiquark channel,
\be
S^{(4)}(x_1, y_1;y_2,x_2)_{\alpha \beta; \delta \gamma} =
<0|T \, q_{\alpha}(x_1) \bar{q}_{\beta}(y_1) q_{\gamma}(x_2)
\bar{q}_{\delta}(y_2)|0>_c  \label{4point}
\ee
(flavor and color indices of the quarks are suppressed), with the
aim to derive an equation for its pion pole contribution. As in flat
space,  the Green function (\ref{4point}) satisfies an inhomogeneous
Bethe-Salpeter equation \cite{nam50},
\be
 S^{(4)}(x_1, y_1;y_2,x_2)_{\alpha \beta; \delta \gamma} =
G_0(x_1, y_1;y_2,x_2)_{\alpha \beta; \delta \gamma} \nonumber  +
\int d \mu(z_1) \int d \mu(z_2) \int d \mu(z_3) \int d \mu(z_4)
\\  \times \,\, G_0(x_1, y_1; z_2, z_1)_{\alpha \beta; \alpha'
\beta'} K(z_1, z_2;z_3,z_4)_{\alpha' \beta'; \delta' \gamma'}
S^{(4)}(z_3, z_4;y_2,x_2)_{;\delta' \gamma' \delta \gamma}
\label{inhomBS}
\ee
(the integration measure $d \mu(z)$ was defined in eq.
(\ref{measure})), which can be derived either by summing the
rescattering series of interactions specified by the $\bar{q}
q$-irreducible vertex kernel $K$, or formally from the generating
functional of the NJL model. In eq. (\ref{inhomBS}) we introduced
the quark-antiquark propagator
\be
G_0(x_1, y_1;y_2,x_2)_{\alpha \beta; \delta \gamma} =
S(x_1, y_2)_{\alpha \delta } S(x_2, y_1)_{\beta \gamma}^T,
\ee
where $S$ is the quark propagator on $S^3 \times R$.

Mesonic bound states appear as poles in eq. (\ref{inhomBS}). In
flat space one finds the corresponding intermediate state wave
functions after a four-dimensional Fourier transform as the pole
residua, which satisfy a homogeneous Bethe-Salpeter equation
\cite{nam50}. In general curved space-times the situation becomes
more involved. Due to the lack of translational invariance (and
consequently the absence of a natural global coordinate system,
as provided by the cartesian coordinates in flat space), the Fourier
transform would have to be replaced by a generalized spectral
transform in accord with the symmetry properties of the
metric\footnote{In general curved space-times questions of
principle regarding the consistent definition of particle and
composite-particle concepts \cite{Itz80} can arise. None of these
affect our metric, however, due to its time-like Killing vector. }.

Fortunately, however, the time direction of our metric is flat
and we can isolate the pole piece by a Fourier transform of the
time dependence only. Accordingly, we write the Bethe-Salpeter
amplitude of the pion as
\be
<0| T\, q_{\alpha}(x_1) \bar{q}_{\beta}(y_1) | \pi^a(p)>  = \frac{
e^{-i \omega_{p} X_1^0} }{ \sqrt{2 \omega_{p} V} } \, \chi^a
(x_1,y_1; p)_{\alpha \beta} \label{BSampl},
\ee
where $p = \{\omega_{\bf p}, {\bf p} \} $ stands for the set of
four quantum numbers which replace the flat-space four-momentum
vector of a free pion. The pion energy $\omega_{\bf p}$
and the conserved ``momentum'' quantum numbers,  ${\bf p} =
\{ n,l,m \} $, specify the spatial state of the pion completely and
are discussed in more detail in appendix D. The notation is intended
to emphasize the similarities with the standard quantum numbers in
flat space. The time coordinates of the center of mass are
\be
X_i^0 = \frac{1}{2} (x_i^0 + y_i^0), \quad \quad i=1,2.
\ee
The time dependence on the right-hand side of eq. (\ref{BSampl}) is
entirely determined by time translation symmetry. For the same
reason
$\chi^a$ depends only on time differences, which we will, however,
not exhibit explicitly. Finally, the normalization of the
Bethe-Salpeter amplitude agrees with the normalization of the
one-pion states adopted in appendix D.

For the course of the following derivation it is convenient to
introduce a finite pion mass, which allows the specialization to
the pion rest frame. Later we will go back to the chiral limit.
The bound state (pion) contribution to $S^{(4)}$, obtained by
inserting on-shell pion intermediate states into eq.
(\ref{4point}), can now be written in terms of the Bethe-Salpeter
amplitude and its conjugate:
\be
i \int \frac{d p_0}{2 \pi} \sum_{{\bf p},a} \frac{ \chi^a
(x_1,y_1; p)_{\alpha \beta}
\bar{\chi}^a (x_2,y_2; p)_{\delta \gamma} }{p_0^2 - \omega_p^2 + i
\epsilon}  \, e^{-i p_0 (X_1^0 - X_2^0)}
\ee
After inserting this contribution into eq. (\ref{inhomBS}) and
specializing the external momentum to the pion mass shell, we end
up with the homogeneous Bethe-Salpeter equation for $\chi^a$:
\be
e^{-i p_0 X_1^0} \chi^a (x_1,y_1; p)_{\alpha \beta} = \int d \mu(z_1)
\int d \mu(z_2) \int d \mu(z_3) \int d \mu(z_4) \nonumber \\ \times
 \,\, G_0(x_1, y_1; z_2, z_1)_{\alpha \beta; \alpha' \beta'}
K(z_1, z_2;z_3,z_4)_{\alpha' \beta'; \delta' \gamma'} \chi^a
(z_3,z_4; p)_{\delta' \gamma'} \, e^{-i p_0 (z_3^0 - z_4^0)/2 }.
\label{homBS}
\ee
In order to proceed further, the interaction kernel $K$ has to be
specified. The standard ladder approximation to the Bethe-Salpeter
equation is consistent with the Hartree-Fock approximation for the
quark propagator from the last section. We thus adopt it here and
choose $K$ accordingly to contain just the tree-level contact
interaction of the NJL model:
\be
K(z_1, z_2;z_3,z_4)_{\alpha \beta; \delta \gamma} = 2 i g
(\Gamma_{\alpha \beta}^A \Gamma_{\gamma \delta}^A - \Gamma_{\alpha
\delta}^A \Gamma_{\gamma \beta}^A) \delta^4(z_1, z_2) \delta^4(z_2,
z_3)
\delta^4(z_3, z_4).
\ee
As a consequence, eq. (\ref{homBS}) reduces (from now on we use
matrix notation in flavor, color and spinor space) to
\be
e^{-i p_0 X^0} \chi^a (x,y; p) = 2 i g \int d \mu(z) \{ tr[\Gamma^A
\chi^a (z,z; p) ] S(x,z) \Gamma^A S(z,y) \nonumber \\ - S(x,z)
\Gamma^A \chi^a (z,z; p) \Gamma^A S(z,y) \} e^{-i p_0 z^0}.
\label{NJLBS}
\ee
In flat space this equation has been solved analytically by Nambu
and Jona-Lasinio in their original paper \cite{NJL69}. The analytical
solution ows its existence to the zero range of the interaction, and
an analogous solution can thus be expected in the curved unit cell.

We will now derive this solution explicitly from an infinitesimal
chiral transformation of the quark propagator. Writing the chiral
generators as $T^a_5 = \frac{i \tau^a}{2} \gamma_5$ and using the
explicit form of the self-energy (\ref{sol}) with the self-consistent
Hartree-Fock propagator, we find
\be
\{ \Sigma (z), T^a_5 \} =  - 2  i g  \left[  \,\Gamma_A \,
{\rm tr} \, [ \Gamma_A \{ S(z,z), T^a_5 \}]  - \Gamma_A \,\{
S (z,z) , T^a_5 \} \, \Gamma_A \,  \right] .  \label{trf}
\ee
This transformation behavior follows directly from the chiral
invariance of the NJL interaction. (Note that the two coincidence
limits $t \rightarrow \pm \infty$  of $S$ differ only in the vector
part (cf. eq. (\ref{coin})). Since the latter drops out of eq.
(\ref{trf}), their further distinction becomes unnecessary.) Now
we can write
\be
\{ S(x,y), T^a_5 \} &=&  \int d \mu(u) \int d \mu(v) S(x,u) \{
S^{-1}(u,v) , T^a_5 \}  S(v,y) \nonumber \\ &=& - \int d \mu(z)
S(x,z) \{ \Sigma (z) , T^a_5 \}  S(z,y) \nonumber \\ &=&  2  i g
\int d \mu(z) S(x,z) \left[  \,\Gamma_A \, {\rm tr} \, [ \Gamma_A
\{ S(z,z), T^a_5 \}]  - \Gamma_A \,\{  S (z,z) , T^a_5 \} \,
\Gamma_A \, \right]  S(z,y), \label{deriv}
\ee
which has exactly the form of the Bethe-Salpeter equation
(\ref{NJLBS}) for zero-momentum pions in the chiral limit, with
the solution
\be
\chi^a (x,y) = {\cal N} \, \{ S(x,y), T^a_5 \} = 2 \, {\cal N}
\,\Sigma_s \int d \mu(z) S(x,z) \, T^a_5 \, S(z,y) \label{piwf}
\ee
for the pion wave function. The normalization ${\cal N}$ will be
fixed in the next section from the chiral Ward-Takahashi identity.
Equation (\ref{piwf}) remains a solution of the Bethe-Salpeter
equation for a small, finite pion mass $m_{\pi}$, up to corrections
of order $m_{\pi}^2$.

As the longest-wavelength excitations of the vacuum, the pions and
their Bethe-Salpeter amplitude can be expected to depend strongly
on the vacuum structure and its variations with density. In the next
section we will study this issue further by calculating the pion
decay constant and its density dependence, building on the results
derived above.

\section{The Pion Decay Constant}

A second key parameter in the discussion of spontaneous chiral
symmetry breaking, besides the quark condensate, is the decay
constant of the pion, $f_{\pi}$. The explicit expression for the
Bethe-Salpeter amplitude from the last section allows us to adapt
the standard method of Nambu and Jona-Lasinio to calculate $f_{\pi}$
in the curved unit cell. We start from its definition in terms
of the axial current matrix element between the one-pion state and
the vacuum,
\be
<0|j_{5,\mu}^a(x)|\pi^b(p)> = - f_{\pi}  \delta^{a b} \partial_{\mu}
\eta_{\bf p} (x). \label{axme}
\ee
The space-time dependence of eq. (\ref{axme}) follows entirely from
the spatial $SO(4)$ symmetry of the hypersphere and from time
translational invariance.
It is contained in the eigenmodes $\eta$ of the Klein-Gordon equation
on $S^3 \times R$, which generalize the usual plane waves of Minkowski
space and are explicitly given and normalized in appendix D, where
also the set of generalized momenta ${\bf p}$ is defined.

By covariantly differentiating eq. (\ref{axme}) one obtains
the hypersphere equivalent of the PCAC relation between the vacuum
and one-pion states,
\be
<0|\nabla^{\mu} j_{5,\mu}^a (x)|\pi^b(p)> &=& - f_{\pi} \,
\delta^{a b} g^{- \frac12} \partial^{\mu} g^{ \frac12} g_{\mu \nu}
\partial^{\nu}  \eta_{\bf p} (x) = f_{\pi} m^2_{\pi} \delta^{a b}
\eta_{\bf p} (x). \label{pcac}
\ee
In the derivation of eq. (\ref{pcac}) we used the free Klein-Gordon
equation (\ref{kgeq}) and the standard definition of the covariant
derivative of a vector field,
\be
\nabla^{\mu} A_{\nu}(x) \equiv \partial^{\mu} A_{\nu}(x) +
\Gamma^{\,\,\,\mu \rho}_{\nu} A_{\rho}(x),
\ee
which contains the Christoffel connection introduced in eq.
(\ref{crist}). The relation to the Laplace-Beltrami operator is
immediately established with the help of the identity $\Gamma^{\,\,\,\nu
\mu}_{\nu} = g^{- \frac12} (\partial^{\mu} g^{ \frac12} )$. We further
introduce the axial-vector and pseudoscalar current densities of
the NJL model,
\be
j_{5,\mu}^a (x) = \bar{q} (x) \gamma_{\mu} \gamma_{5} \frac{\tau^a}{2}
q (x), \qquad \quad j_{5}^a (x) = \bar{q} (x)  i \gamma_{5}
\frac{\tau^a}{2} q (x), \label{curr}
\ee
which are connected by the divergence of the axial current,
\be
\nabla^{\mu} \,j_{5,\mu}^a (x) = 2 m_0 \, j^a_{5} (x). \label{currdiv}
\ee
The relation between the pion decay constant and the Bethe-Salpeter
amplitude, eq. (\ref{BSampl}), is established through the
axial current matrix element (\ref{axme}), which can be expressed
in terms of the coincidence limit of $\chi^a$:
\be
<0|j_{5,\mu}^a (x)|\pi^b(p)> = - \frac{ e^{-i
\omega_{\bf p} X^0} }{\sqrt{2 \omega_{\bf p} V} } \, {\rm tr}
[ \gamma_{\mu} \gamma_{5} \frac{\tau^a}{2} \chi^b (x,x; p) ].
\label{axme2}
\ee
(The trace is over Dirac, flavor and color indices.)

Before calculating $f_\pi$ from this relation, we have to fix the
normalization of the pion wave function. To this end we consider
the quark-antiquark three-point functions of the current densities
(\ref{curr}),
\be
G^a_{5, \mu} (z,x,y) = <0|T\, j^a_{5,\mu} (z) q(x) \bar{q} (y) |0> ,
\ee
\be
G^a_{5} (z,x,y) = <0|T\, j^a_{5} (z) q(x) \bar{q} (y) |0>,
\ee
and derive with the help of eq. (\ref{currdiv}) the chiral
Ward-Takahashi identity
\be
\nabla_{(z)}^{\mu} \, G^a_{5, \mu} (z,x,y) =  2 m_0 \, G^a_{5}
(z,x,y) -   T^a_5 \, S(z,y) \, \delta^4 (x,z)  -  S(x,z) \,
\delta^4 (z,y) \, T^a_5.
\ee
Integrating over $z$, the surface term on the left-hand side
vanishes (recall that the pions have a finite mass) and we obtain
\be
2 m_0 \, \int d \mu(z) \, G^a_{5} (z,x,y) =  \{ S(x,y), T^a_5 \}.
\label{chiwi}
\ee
The left-hand side receives contributions from one-pion
intermediate states, which can be expressed in terms of the
Bethe-Salpeter amplitude as
\be
2 m_0 \, \int d \mu(z)  \, G^a_{5} (z,x,y) =  -i f_{\pi} \chi^a
(x,y), \label{bs2}
\ee
and, comparing eq. (\ref{piwf}) with eqs. (\ref{chiwi}) and
(\ref{bs2}), we read off the normalization constant
\be
{\cal N} = \frac{ i}{f_{\pi}}.
\ee

With the normalization fixed, we can now equate the time components
of the two expressions for the axial current matrix element, eqs.
(\ref{axme}) and (\ref{axme2}), in the pion rest frame ${\bf p} =
\{0,0,0\}$ with (cf. appendix D)
\be
\eta_{ \{0,0,0\} }(x) = \frac{e^{- i m_{\pi} t}}{\sqrt{2 m_{\pi} V}},
\ee
to obtain an equation for the pion decay constant in terms of the
Bethe-Salpeter amplitude:
\be
i m_{\pi} f_{\pi} = - {\rm tr} \, \left[ \gamma_{0} \gamma_{5}
\frac{\tau^3}{2} \chi^3 (x,x; p = m_{\pi}) \right].
\ee
After inserting the explicit solution in the pion rest frame, eq.
(\ref{piwf}), we have
\be
i m_{\pi} f^2_{\pi} = \Sigma_s \int d \mu(x) e^{- i m_{\pi} t}
{\rm tr}  \, \left[  \gamma_{0} \gamma_{5} \frac{\tau^3}{2} S(0,x)
\gamma_{5} \tau^3  S(x,0)\right]. \label{fpieq1}
\ee
We now use the Hartree-Fock Dirac propagator on $S^3 \times R$
from section III to evaluate the right hand side further. The color
and flavor traces in eq. (\ref{fpieq1}) produce a trivial factor
$n_c n_f$ and the complete trace (over color, flavor and Dirac
indices) becomes
\be
{\rm tr} [  \gamma_{0} \gamma_{5} \frac{\tau^3}{2} S(0,x) \gamma_{5}
\tau^3  S(x,0)] &=&  2 n_c n_f  \left[ S_0(\mu, -t) S_2(\mu, t) -
S_0(\mu, t) S_2(\mu, -t) \right] \nonumber \\ &=& - 4 n_c n_f
S_0(\mu, t) S_2(\mu, t) = \frac{2 i}{m} \frac{\partial}{\partial t}
S_2^2(\mu, t)
\ee
($m$ is the dynamical quark mass), where we made use of the
time-reversal properties of the propagator functions,
\be
S_0(\mu, -t) = - S_0(\mu, t), \qquad S_2(\mu, -t) = S_2(\mu, t),
\ee
and of the identity
\be
S_0(\mu, t) = \frac{- i}{m} \frac{\partial}{\partial t}  S_2(\mu, t),
\ee
which follows directly from the explicit expressions for $S_0$ and
$S_2$ in appendix B. After performing the two trivial angular
integrations in the measure eq. (\ref{measure}),
and a partial integration in the time coordinate, the equation for
the decay constant simplifies to
\be
f^2_{\pi} = - 4 i \pi  n_c n_f L^3 \int_{-\infty}^{\infty} dt
e^{- i m_{\pi} t} \int_0^{2 \pi} d \mu \, \sin^2 \mu \,\, S_2^2
(\mu, t). \label{fpieq2}
\ee
We now use the spectral representation of the propagator function
$S_2$, eq. (\ref{S2spec}), to write
\be
\int_{-\infty}^{\infty} &dt& e^{- i m_{\pi} t} \int_0^{2 \pi}
d \mu \, \sin^2 \mu
\,\, S_2^2 (\mu, t)   \\ & & = \left( \frac{ i m}{4 V} \right)^2
\sum_{n,m =1}^{ \infty}  \int_0^{2 \pi} d \mu  \,\, s_n(\mu)\,
s_m(\mu)  \int_{-\infty}^{\infty} dt \frac{ {\rm e}^{-( i \omega_n
+ i \omega_m + \epsilon) |t| }}{\omega_n \omega_m} ,
\ee
where
\be
s_n(\mu) =  2 k_n L \sin (k_n L \mu) - \cos (k_n L \mu) \tan
\frac{\mu}{2}.
\ee
The  integral over $\mu$ can be done explicitly,
\be
\int_0^{2 \pi} d \mu  \,\, s_n(\mu) \, s_m(\mu) = 2 \pi \delta_{n m}
D_n,
\ee
with the quark level degeneracies $D_n$ given in eq. (\ref{degen1}).
It remains to combine the above equations, to regularize the remaining
mode sum consistently with the gap equation (by using the same regulator,
eq. (\ref{cutoff})) and to introduce the filling factors of section IV
to account for the valence quarks. Our final expression for the pion
decay constant then becomes
\be
f^2_{\pi} =  \frac{3 m^2}{2 V}  \sum_{n =1}^{ \infty}
\frac{D_n (1 - f_n)}{\omega^3_n} \Theta_{\epsilon} (\Lambda - k_n).
\label{decay}
\ee

A comparison of this result with the flat-space expression in
ref. \cite{ber87} shows similarities analogous to those between
the expressions for the quark condensate. In particular, $f_\pi$
gets contributions from the quark Dirac sea with the same
functional dependence as in \cite{ber87}. The main difference to
the flat-space result is again due to the new quark spectrum,
which replaces the plane wave spectrum of euclidean space.

We will continue the discussion of eq. (\ref{decay}) and its
density dependence in section VIII.  There we will also present
results from the numerical evaluation and a comparison with the
corresponding results of the chemical-potential approach.

\section{The zero-density limit}

At low baryon densities the volume of the unit cell becomes large
and its curvature goes to zero. In the low-density limit, or
equivalently in the large-$L$ limit, one therefore expects to recover
the results of the flat-space NJL model. In the present section we
will perform this limit explicitly, both as a consistency check on
our results and to illustrate how the quark spectrum and its
degeneracies turn into the plane wave form.

Let us start by considering the $L \rightarrow \infty$ limit of the
scalar self-energy, eq. (\ref{sselffinal}). In this limit the Fermi
momentum goes to zero\footnote{Recall that the Fermi momentum sets
the
baryon number in the cell to one. } and contributions from quark
propagation around the whole sphere are suppressed. Therefore it is
convenient to use the representation of the propagator in terms of
a sum over paths. Up to contributions from complete circles around
the hypersphere we obtain
\be
\Sigma_s &=&  \frac{- g m}{4 \pi^2 L^2} (1+2 n_c n_f)
\int_{k_F}^{\infty} dk \frac{1 - 4 k^2 L^2}{\omega} \Theta_{\epsilon}
(\Lambda - k)  \\
&\rightarrow& \frac{g m}{ \pi^2 } (1+2 n_c n_f) \int_0^{\infty} dk
\, \frac{k^2}{\omega} \, \Theta_{\epsilon} (\Lambda - k),
\ee
which is indeed the original flat space result \cite{NJL69}.
Similarly,
the vector part of the self-energy becomes
\be
\Sigma_v &=& \frac{ g}{2\pi^2 L^2}  \int_0^{k_F} dk (1 - 4 k^2
L^2) \\
&\rightarrow&  0,
\ee
and goes, as expected, to zero in the large-$L$ limit.

To check the zero-density limit of $f^2_{\pi}$ is a bit more subtle,
since the corresponding loop contains two quark propagators instead
of the coincidence limit of only one. We start from eq. (\ref{fpieq2})
and use the representation (\ref{S2}) for the propagator function
$S_2$. Neglecting again additional complete turns around the
hypersphere, we have
\be
f^2_{\pi} &=& - 4 i \pi  n_c n_f (\alpha m)^2 L^3 \int_0^{\infty}
dt  \int_0^{2 \pi} d \mu [2 I_0'(\mu,t) + \tan \frac{\mu}{2}
I_0(\mu,t) ]^2 \nonumber \\
&=& - \frac{i n_c n_f m^2}{4 \pi L} \int \frac{dk_0}{2 \pi}
\int_{-\Lambda}^{\Lambda} \frac{dk}{2 \pi} \int_{-\Lambda}^{\Lambda}
\frac{dq}{2 \pi}  \int_0^{2 \pi} d \mu \frac{e^{iL(k+q) \mu}
\left[ 2ikL + \tan \frac{\mu}{2} \right]  \left[ 2iqL + \tan
\frac{\mu}{2} \right] }{[k_0^2 - k^2 -m^2 + i \epsilon ] [k_0^2 -
q^2 -m^2 + i \epsilon ] }.
\ee
In the second line the time integration, which leads to a delta
function  $\delta (k_o -q_0)$, and the consecutive $q_0$ integration
are already performed. Note that the range of the $\mu$ integration
in the above expression can be changed to $[- 2 \pi, 0]$ without
changing the integral, if we simultaneously change the signs of the
integration variables $k$ and $q$. We can thus replace half of
the $\mu$-integral above by an integral with the alternative,
equivalent
range. After furthermore scaling $\mu$ to $\tilde{\mu} = \mu L$,
the
$L$-dependent part of the above expression becomes
\be
\frac{1}{2 L} \int_{-\Lambda}^{\Lambda} \frac{dk}{2 \pi}
\int_{-\Lambda}^{\Lambda} \frac{dq}{2 \pi}  \int_{-2 \pi L}^{2 \pi L}
d \tilde{\mu} e^{i(k+q) \tilde{\mu} } \left[ 2ikL + \tan
\frac{\tilde{\mu}}{2} \right]  \left[ 2iqL + \tan
\frac{\tilde{\mu} }{2} \right],
\ee
which, in the large-$L$ limit, reduces to
\be
\int_{-\Lambda}^{\Lambda} \frac{dk}{2 \pi} \int_{-\Lambda}^{\Lambda}
\frac{dq}{2 \pi} \frac{\pi}{L} \delta (k+q) [-4 k q L^2 ].
\ee
Inserting this in the expression for $f^2_{\pi}$ above, we finally
get
\be
f^2_{\pi} &\rightarrow& -i n_c n_f m^2 \int_{-\Lambda}^{\Lambda}
\frac{dk}{2 \pi} k^2 \int \frac{dk_0}{2 \pi} \frac{1}{[k_0^2 - k^2 -
m^2 + i \epsilon ] [k_0^2 - k^2 -m^2 + i \epsilon ]  } \nonumber \\
&=& \frac{n_c n_f m^2}{4 \pi^2} \int_0^{\Lambda} dk
\frac{k^2}{\omega^3},
\ee
($\omega = \sqrt{k^2 + m^2}$), which agrees with the flat-space
result
\cite{NJL69}.

The large--$L$ limit thus reproduces, as expected, the standard
zero--density quantities of the NJL model. The transition of the
physics from the high-- to the low--density regime can be traced
to the differences in the quark propagation in small and large
hyperspheres. At high densities geodesic paths with any number of
turns around the sphere contribute and their sum builds up to the
equivalent sums over the quark spectrum on $S^3 \times R$ (cf.
appendix C). In the large-$L$ limit, on the other hand, only the
direct path is relevant and yields the standard flat-space
propagator.

\section{Results \label{results} }

Before embarking on the quantitative discussion of our results, we
fix the coupling $g$ and the cutoff $\Lambda$ of the NJL model at
$g= 4.08 \, {\rm GeV}^{-2}$ and $\Lambda = 700 \, {\rm MeV}$.
These values have been found in ref. \cite{ber86} to best reproduce
the phenomenological values of the quark condensate and of the pion
decay constant in the vacuum. We adopt this choice to allow for a
direct comparison between our results and those of ref. \cite{ber87},
and to ensure a phenomenologically acceptable zero-density limit of
both the quark condensate and of $f_{\pi}$. Nevertheless, given the
simplicity of the minimal NJL model and the sensitivity to the values
of $\Lambda$ and $g$ already noted in ref. \cite{ber87}, our numerical
results should be considered as qualitative.

The above model parameters could in principle be density dependent.
Since we are mainly interested in new, qualitative features of the
hypersphere formulation, however, we will follow the standard
treatment and keep them constant. Finally, the
diffuseness parameter in the regulator, eq. (\ref{cutoff}), is
set to $\epsilon = 2.4$. This choice compromises between a rapid
convergence of the mode sums and a sufficiently smooth behavior of
the observables at high densities.

We are now ready to discuss our numerical results. The non-trivial
solution of the gap equation (\ref{gapeq}), {\it i.e.} the dynamical
quark mass $m$, and the negative cube root of the corresponding
quark  condensate, $-<\bar{q} q>^{1/3}$, are shown as a function
of density
in fig. 1. Both quantities decrease monotonically with density and
finally go to zero at the density $\rho_c$.

The vanishing quark condensate does of course not necessarily imply
a {\it second-order} phase transition to chiral restoration. Recall
that the condensate shown in fig.1 was
calculated by using the nontrivial solution of the gap equation
in the
quark propagator. It is not {\it a priori} clear, however, that
this
solution remains physical, {\it i.e.} the one of lowest free
energy, if the baryon density increases.

Indeed, it was claimed in ref. \cite{asa89} that in a certain
parameter range of the standard NJL model, close to the chiral
limit, the trivial solution becomes the global minimum of the free
energy {\it before} the nontrivial solution reaches zero. This result,
obtained in the chemical-potential approach, would imply a first
order
phase transition to chiral restoration. However, studies in more
realistic versions of the NJL model found a second-order transition
for the whole phenomenologically acceptable parameter range
\cite{kli90}.

To determine the order of the phase transition in our approach, we
monitor the extrema of the free energy density
\be
\frac{\Omega(V, m)}{V} = \frac{g n_f}{n_c} (1 + 2 n_c n_f)
<\bar{q} q>^2 - \frac{g n_f}{n_c} <q^{\dagger} q>^2 -
\frac{n_c n_f}{V} \sum_{1}^{\infty}
D_n (1 - f_n) \omega_n  \Theta_{\epsilon} (\Lambda - k_n)
\label{fren}
\ee
as a function of the cell volume or, equivalently, of the baryon
density. The last term in eq. (\ref{fren}) is just the regulated
sum over the energies of the occupied quark states, while the first
two terms are required to avoid double-counting of the vacuum
energy \cite{suz63}. It is straightforward to check that the
extrema of eq.
(\ref{fren}) under variations of the quark mass are, as expected,
the solutions of the gap equation (\ref{gapeq}).

The dependence of $\Omega/ V$ on the quark mass is plotted in
fig. 2 for four baryon densities between $\rho = 0$ and $\rho =
\rho_c$. These curves demonstrate that only two solutions of the
gap equation exist at all densities up to $\rho_c$: the trivial
one at $m=0$ and the nontrivial one with finite $m$. In particular,
the nontrivial (trivial) solution remains the minimum (maximum) of
the free energy density up to $\rho_c$, where both extrema merge.
The phase transition is therefore of second order.

Closer inspection of the free energy density reveals that this
result is, in contrast to the chemical potential approach,
independent of the model parameters. Indeed, it is straightforward
to show that, as long as the nontrivial solution exists, it remains
the minimum of the free energy density. This robust second-order
phase transition, which actually persists
for a much larger class of NJL models, is a welcome result, since
the specific choice of the lagrangian is somewhat arbitrary. It
also insures that a potential density dependence of the model
parameters would not affect the order of the phase transition.

For comparison, the quark condensate as given in ref. \cite{ber87},
{\it i.e.} calculated on the basis of the nontrivial solution in
the chemical-potential approach, is also shown in fig. 1. There may
exist parameter regions in which the transition of ref. \cite{ber87}
becomes actually first order in the chiral limit. Introducing a
current quark mass would in this case restore the second
order transition without significant effects on the curve
and the approximate critical density. In order to compare
the density dependence and the critical densities of the
{\it second-order} transitions, we therefore reproduce the
original plots of ref. \cite{ber87} in fig. 1.

Both approaches yield a very similar behavior of the order
parameter with density. This is reminiscent of the analogous
situation in the Skyrme model, where close similarities between
hypersphere and array results first demonstrated the use of
hypersphere calculations.

In fig. 3 we plot the density dependence of the pion decay constant
and again, for comparison, the same quantity as obtained in the
chemical potential approach. $f_{\pi}$ decreases with density and
goes to zero at the chiral restoration density. This is expected,
since the pion has to decouple when the chiral condensate disappears.
In the hyperspherical cell $f_{\pi}$ decreases  initially somewhat
faster then in the chemical potential approach, but goes to zero at
almost the same critical density.

Figure 4 shows the density dependence of both the quark condensate
and the pion decay constant in the {\it absence} of valence quarks
in the cell. This situation corresponds to setting $k_F = 0$ or,
equivalently, to setting all $f_n = 0$. Even in this case both the
condensate and the decay constant decrease and eventually vanish.
This happens, however, at a considerably higher density.
The absence of baryon number sources inside the cell delays the
transition to chiral restoration, as one would expect.

The valence-quark free cell can be interpreted as a region of
reduced baryon density in the nuclear medium. While the
interaction with the surrounding matter is still mediated by
the shape and curvature of the $B=0$ cell, it is felt only by
the quark vacuum in the interior. Such $B=0$ cavities can therefore
neither be studied in the chemical-potential approach to the
NJL model nor in the Skyrme model. In the latter, a unit of baryon
number in {\it every} cell is crucial for the chirally restored
phase to exist, at least in semiclassical approximation. In the
standard approach to the NJL model at finite density, on the other
hand, a {\it homogeneous} baryon density distribution is built in
from the beginning.

Due to the fixed relation between the ambient baryon density and the
radius of {\it all} cells, however, the size of the $B=0$ cavity
cannot be varied at a given density. Such baryon-density free bubbles
with delayed chiral restoration will therefore very likely not
describe
an equilibrium situation, but they might be related to a transient
bubble phase in heavy-ion reactions, where the chirally broken and
restored phases coexist.

\newpage

\section{Summary and Conclusions}

In this paper we have generalized a new approach to dense nuclear
matter in the framework of extended hadron models. Originally
developed for the Skyrme model, it divides the nuclear medium into
cells of unit baryon number, which have the form of a 3-dimensional
hypersphere $S^3$. The new and crucial feature of these cells is
their intrinsic curvature, which mediates interactions with the
ambient matter.

In order to explore the use of this approach beyond the Skyrme model,
we first extend it to quark-based baryon models in general and then
apply it specifically to the Nambu-Jona-Lasinio model, by investigating
vacuum, constituent-quark and pion properties as well as their density
dependence. The presence of fermions in the curved cell requires the
introduction of additional concepts, under which a new, isoscalar gauge
interaction for the quarks is the most important.

We find a high-density phase transition, which restores the chiral
symmetry of the vacuum by decondensing quark-antiquark pairs. At the
same critical density the pions decouple, as signaled by the vanishing
of the pion decay constant, which monotonically decreases with density.
The phase transition is, for a large class of generalized NJL models,
robustly of second order and independent of the model parameters.
Furthermore, delayed chiral restoration takes place even inside of
valence--quark--free cavities in the dense medium. In this case it is
solely driven by the interaction of the quark vacuum with the cell
curvature.

Remarkably, we find close similarities between the density dependence
of our results and those of the conventional chemical-potential
approach. The behavior of the constituent quark mass and of the quark
condensate with density, for example, is (in the  phenomenologically
reasonable NJL parameter range) in both cases almost identical. This
similarity may seem at first surprising, since the two approaches are
based on rather different descriptions of the dense environment. At
least qualitatively, however, it is easy to understand. Recall that
the energy gain from the Dirac sea, filled with massive as
compared to massless quarks, leads at zero baryon density and
for sufficiently large coupling to the development of a nontrivial
vacuum. In the chemical potential approach, this energy gain gets
with increasing density more and more compensated by the occupied
states in the Fermi sea, since they count with opposite sign in the
spectral sum of the gap equation. At the critical density, it
becomes energetically favorable for the quarks to be massless,
and the chiral phase transition occurs.

On the hypersphere the mechanism is different in principle, but
similar in effect. The occupation numbers of the valence quark levels
are here density-independent, but the {\it spectrum itself} changes
in a characteristic way. Due to the stronger localization in the
smaller high--density cells, the quark momenta grow inversely with
the cell size. This leads to a reduced number of states in the
regularized Dirac sea, with the same consequences as in the
chemical potential approach.

The overall increase of the quark momenta at larger densities is
an effect common to all compact unit cells. The specific and
detailed change of the quark spectrum, however, and therefore the
quantitative density dependence of the results, is governed by
the particular form of the unit cell. The close agreement with the
chemical-potential approach thus supports the choice of $S^3$ as
the cell geometry. An analogous agreement between results of the
conventional and the hypersphere approach was observed in the
Skyrme model.

Experience gained from hypersphere calculations in different
models can help to disentangle features specific to the Skyrme
model from more general or even model-independent aspects of the
hypersphere approach. In this respect our results indicate,
for example, that the presence of a soliton is not required for
the hypersphere approach to be effective. In
particular, the specific topological properties of the Skyrme
soliton and its hedgehog form are not indispensable. Since the
winding number of the skyrmion plays an important role in its
hypersphere behavior, this was not {\it a priori} obvious.
Furthermore, a non-zero baryon number ({\it i.e.} winding number)
in the cell and the corresponding hedgehog structure are necessary
requirements for chiral restoration in the Skyrme model, whereas
this
condition is suspended in the NJL model, as the results in the
$B=0$ cell show. The NJL results also indicate that the hypersphere
approach is compatible with different mechanisms for spontaneous
chiral symmetry breaking.

The application of the hypersphere approach has often practical
advantages, which may, however, vary in different models. In the
Skyrme model, for example, $S^3$ calculations require drastically
less numerical effort than conventional unit-cell calculations.
While this particular benefit does not translate to the NJL case
(since the numerical requirements of the conventional
(chemical--potential) approach are already moderate), the hypersphere
approach to the NJL model can be applied in realms not easily accessible
by other methods. The discussed low-density bubbles in nuclear matter
with delayed chiral restoration can {\it e.g.} neither be studied
in the Skyrme model nor in the chemical-potential approach to the
NJL model. Another profitable application of the hypersphere approach
to the NJL model is encountered in the study of nuclear matter at
finite temperature. While the Skyrme model calculation meets with
rather involved technical difficulties in this situation even on
the hypersphere, it can be straightforwardly addressed in the NJL
case
\cite{for93}.

We regard the above results as promising indications for the
hypersphere approach to be useful beyond the Skyrme model.
Nevertheless, many important and interesting questions, also on a
more fundamental level, remain open. It would, for example, be
useful
to know if hyperspherical and flat unit cells can be explicitly
related to each other, at least approximately. A hint in this
direction comes from the
work of ref. \cite{jac87}, where matter at a fixed baryon density
is almost identically described in hyperspherical cells with both
baryon number and volume doubled, as a first step towards the limit
of a skyrmion matter configuration in flat space. One might also
hope to make contact with older speculations \cite{pak79} about a
description of chiral dynamics in terms of space curvature.

The theoretical framework developed in the present paper can
straightforwardly be adapted to other chirally symmetric quark and
hybrid models, as for example to bag models \cite{wei92} and to
non-topological soliton models \cite{fri77}. It might also be
interesting to study the relations between the Skyrme and NJL models,
which can be established by bosonization of the latter \cite{bos,Rei85},
in the hypersphere formulation. Finally, our work gives an explicit
example for symmetry breaking and restoration due to space-time
curvature in a fermionic theory, and as such may be also useful in
other areas of physics. Related questions are, for example, presently
discussed in a cosmological context \cite{det93}.

{\bf Acknowledgements}

It is a pleasure to thank Manoj Banerjee, Jim Griffin, Andy Jackson,
Christian Weiss and Ismail Zahed for useful discussions and comments.
The support of the U.S. Department of Energy under grant no.
DE-FG02-93ER-40762 is also acknowledged.

\newpage

\appendix
\section{Maurer-Cartan Basis and Dirac Spectrum on $S^3 \times R$}

The generalization of Lorentz (Poincar{\'e}) tensor fields to curved
space-times is straightforward. One simply replaces the Lorentz group
by the group of general coordinate transformations, which becomes
locally (\ie in the tangent spaces of the manifold) the full linear
group $GL(R,4)$. The generalized tensor fields form its representations.

The situation is more involved for fields of half-integer spin,
because the linear group has no double-valued ({\it i.e.} spinor)
representations. For this reason, spinors have to be introduced with
respect to a local, orthonormal basis, the vierbein $e_a$ (in the
(co)tangent space of space-time). As discussed in the introduction,
the metric and therefore the physics is independent of the arbitrary
orientation of these frames, which can be locally changed by a
Lorentz gauge group. The fermion fields form
the spinor representations of this local Lorentz group, which acts in
its defining representation on the Latin index of the
vielbein\footnote{This {\it local} definition of the spinors can in
general not be consistently extended to the manifold as a whole.
$S^3 \times R$, however, as a group manifold, is parallelizable.
In technical terms this implies the vanishing of its first two
Stiefel-Whitney classes, which guarantees the existence of a {\it
global} spin structure.}.

For explicit calculations we have to choose an appropriate vielbein
on $S^3 \times R$, \ie we have to fix the gauge of the local Lorentz
group. The spatial hypersphere has the geometry of the group manifold
of $SU(2)$, scaled to radius $L$. This suggests the rescaled
Maurer-Cartan vector fields of SU(2) as the natural choice for the
spatial components of the vielbein\footnote{To be specific, we use
the right-invariant Maurer-Cartan fields. This is purely a matter of
convention.}
\be
e_a = \frac{L}{2i} {\rm Tr} [\, \tau_a(\partial_i h)h^{-1} \,]\,\,
 g^{ij} \,\,\del_j  \qquad
(a,i =1,2,3),
\ee
where we parametrized the elements of $SU(2)$ in polar coordinates as
\be
h(x) = \cos \mu + i\tau_a{\hat r}^a(\theta,\phi) \sin \mu.
\ee
(${\hat r}^a(\theta,\phi)$ is the unit vector in $R^3$.) Complemented
by the time component $e_o=\del_t$ we obtain an orthonormal vierbein
on
$S^3(L) \times R$, from which the metric eq. (\ref{metric}) can be
recovered with the help of eq. (\ref{metviel}). The simplest way to
derive the corresponding Levi-Civita spin connection is to solve
Cartan's first structure equation \cite{Egu77} in the same
``Maurer-Cartan gauge''\footnote{A fixed vierbein does in general not
determine the corresponding spin connection completely. We take our
cell to be torsion-free, however, and further require the spin
connection to be compatible with the metric (metricity condition),
so that it is {\it uniquely} given by the first Cartan structure
equation.}. The result is
\be
\omega_a = \frac{i}{4L} \epsilon_{abc} \sigma^{bc} \quad (a,b,c
\,\,\epsilon \,
\{ 1,2,3 \} ) \qquad {\rm and} \quad \omega_o = 0.
\ee
($\sigma_{ab} = \frac{i}{2} [\gamma_a,\gamma_b]$) Note that the
$\omega_a$ are space-time independent and essentially given by
the structure constants of $SU(2)$, which is the main advantage of
using the Maurer-Cartan vielbein.

The Dirac operator on $S^3 \times R$, which appears in the free part
of the NJL lagrangian, eq. (\ref{njlhyp}), and to which we add a
constant vector self-energy (we do not include the scalar part of the
self-energy, since it will later be absorbed in the mass term), is
\be
 i \slash{\mkern - 3mu D} = [ i \gamma^a ( e_a + \omega_a) +
\Sigma_v \gamma_0 ] \label{diracop}
\ee
and now takes the simple form\footnote{We use the chiral
representation of the $\gamma$-matrices: $\gamma^o \equiv \beta =
\left( \begin{array}{cc} 0 & -I \\ -I  & 0 \end{array} \right),
\vec{\gamma} = \left( \begin{array}{cc}  0 & \vec{\sigma} \\ -
\vec{\sigma} & 0 \end{array} \right), \vec{\alpha} = \left(
\begin{array}{cc} \vec{\sigma} &  0  \\ 0 & -\vec{\sigma}
\end{array}
\right), \gamma_5 = \left( \begin{array}{cc} I & 0 \\ 0  & -I
\end{array}
\right)$}
\be
 i \slash{\mkern - 3mu D} = \left( \begin{array}{cc} 0
& \Delta_{+} \\ \Delta_{-} & 0 \end{array} \right),
\label{blockop}
\ee
in terms of the operators
\be
\Delta_{\pm} = -  i \del_t - \Sigma_v \pm ( i \vec{\sigma}
\vec{e} + \frac{3}{2L}). \label{deltapm}
\ee
(Arrows indicate vectors in the spatial, euclidean subspace of the
local vielbein.)

For the interpretation of our results it will be useful to derive
the energy spectrum of the corresponding Dirac equation. Let us
therefore look for the static solutions
\be
\psi_n ({\bf x},t) = \psi_n ({\bf x}) {\rm e}^{- i
\tilde{\omega}_n t}
\ee
of the Dirac equation
\be
( i \slash{\mkern - 3mu D} - m) \psi_n = \left(
\begin{array}{cc} -m
& \Delta_{+} \\ \Delta_{-} & -m \end{array} \right) \left(
\begin{array}{c} \psi^{(1)} \\ \psi^{(2)} \end{array} \right) = 0,
\ee
which has the equivalent hamiltonian form
\be
\left[ -i \vec{\alpha} (  \vec{e} - \vec{\omega} ) + \beta m \right]
\psi_n ({\bf x}) = \omega_n \, \psi_n ({\bf x}),
\ee
where $\omega_n \equiv \tilde{\omega}_n + \Sigma_v$. (The constant
vector self--energy just shifts the origin of the energy scale.) Now
we can again exploit the properties of $S^3(1)$ as the group
manifold of $SU(2)$. The invariant vector fields of $S^3(L)$ are
(as for every Lie group) simply the group generators $\hat{L}_a$,
rescaled to radius $L$:
\be
e_a = \frac{2}{ i L} \hat{L}_a, \label{gen1}
\ee
where the $\hat{L}_a$ satisfy the usual angular momentum
commutation relations
\be
[\hat{L}_a, \hat{L}_b] =  i \epsilon_{a b c} \hat{L}_c.
\label{gen2}
\ee
The appearance of a ``spin-orbit'' term in
\be
\Delta_{\pm} \psi^{a} = \{ - \tilde{\omega} - \Sigma_v \pm
\frac{1}{L} (4 \vec{S} \vec{\hat{L}} + \frac{3}{2} )\} \psi^a \quad
\quad a=1,2
\ee
($\vec{S}$ is the spin operator $\vec{\sigma}/2$) then suggests to
diagonalize $\Delta_{\pm}$ by taking $\psi^a$ in the coupled basis
with $\vec{J} = \vec{\hat{L}} + \vec{S}$, so that
\be
\vec{S} \vec{\hat{L}} \, \psi_{j,m}^a &=& \frac{1}{2} [ j(j+1) -
l(l+1)
- \frac{3}{4}] \, \psi_{j,m}^a \nonumber \\ &=&  \left\{
\begin{array}{ccc} \quad \quad \quad \frac{1}{2} l & \psi_{j,m}^a
& {\rm for} \quad \quad j=l + \frac{1}{2}, \qquad l = \{0,1,2...\}
\\ - \frac{1}{2}(l+1) & \psi_{j,m}^a & {\rm for} \quad\quad j=l -
\frac{1}{2}, \qquad l = \{1,2,3...\}. \end{array} \right.
\label{spinorbit}
\ee
Inserting eq. (\ref{spinorbit}) into the Dirac equation and
combining
the resulting two coupled equations for the upper and lower
spinor
components one obtains
\be
\omega_n^2 = m^2 + \frac{1}{L^2} \times \left\{ \begin{array}{cc}
(2l + \frac{3}{2})^2 & {\rm for} \quad j=l + \frac{1}{2}\\ (-2l -
\frac{1}{2})^2 & {\rm for} \quad j=l - \frac{1}{2} \end{array}
\right. \label{spectrum1}
\ee
which determines the energy eigenvalues $\omega_n$. All properties
of the spectrum can be read off directly from eq. (\ref{spectrum1}):
each pair of levels with $j = \frac{n}{2}, l = \frac{n-1}{2}$ and
$j = \frac{n-1}{2}, l = \frac{n}{2}$, $n= \{1,2,...\}$ contains
\be
D_n = 2n(n+1) \label{degen}
\ee
degenerate states with momentum $k_n$ and energy $\omega_n$,
\be
k_n =  \frac{2n+1}{2L}, \qquad \quad \omega_n^2 = k_n^2 + m^2.
\label{mom}
\ee
Finally, the total number of states up to the $N$th level sums up to
\be
\sum_1^N D_n = \frac{2}{3} N(N^2 + 3 N +2).
\ee
This spectrum agrees with existing derivations in the literature,
obtained by various different methods, see $\eg$ \cite{kov90} and
also \cite{for76} for the massless case. Let us finally mention that
the existence of a time-like Killing vector ($e_0 = \del_t$) in our
cell permits a unique definition of in- and out-states (and their
Fock spaces) on the basis of this spectrum.

\section{The Fermion Propagator on $S^3 \times R$}

\vskip .2 cm

In this appendix we construct the quark propagator $S(x)$ on
$S^3 \times R$ for a Dirac operator of the form discussed in
appendix A, \ie including a constant self-energy (\ref{selfen}).
$S(x)$ is the solution of
\be
( i \slash{\mkern - 3mu D} - m) S(x,y) = \delta^4 (x,y),
\label{propeq}
\ee
with $i \slash{\mkern - 3mu D}$ given in eq. (\ref{diracop}).  The
scalar part of $\Sigma$ is combined  with the current quark mass in
the mass term $m = m_0 + \Sigma_s$, the vector part is absorbed in
the covariant derivative and Feynman boundary-conditions are implied.
(We will later generalize the boundary conditions to take the presence
of valence quarks into account.)

The four-dimensional delta function is generalized to curved space by
demanding
\be
 \int d \mu (x) \, \delta^4 (x,y) f(x) = f(y),
\ee
so that
\be
\delta^4 (x,y) \equiv (- {\rm g})^{-1/2} \prod_{\mu= 0}^3
\delta(x^{\mu} - y^{\mu}),
\ee
up to the sum over periodic paths (see below).  (${\rm g} = -L^6
\sin^4 \mu \sin^2 \theta$ is the determinant of the metric.)

Due to the compactness of the unit cell, the free propagation of a
quark between $x$ and $y$ can proceed either directly (on a geodesic)
or via an arbitrary number of intermediate turns around
the hypersphere. We can therefore consider the geodesic coordinates
as non-periodic real numbers and count the number of windings of a
given path in multiples of $2 \pi$. The propagator then becomes a sum
of the propagators for each of these paths. (In order not to complicate
the following equations unnecessarily, we will suppress these periodic
sums until they play an active role in the derivation.)
The above procedure ensures that the full propagator is periodic
in the
coordinates, as required by the physical situation, and allows to
write its Fourier transform in terms of continuous momenta. Although
the Fourier integrals can be transformed into equivalent sums over
the discrete quark spectrum in the compact space (see appendix C),
the former representation turns out to be more convenient for the
derivation of the propagator.

To derive the Dirac propagator explicitly, we start from the
representation
\be
S(x,y) = - ( i \slash{\mkern - 3mu D} + m) \, G(x,y),
\label{itprop}
\ee
which defines the block-diagonal $4 \times 4$ matrix
\be
G(x,y) = \left( \begin{array}{cc} \bar{G}(x,y)
& 0 \\ 0 & \bar{G}(x,y)  \end{array} \right),
\label{block}
\ee
a Greens function of the iterated Dirac operator. By inserting eq.
(\ref{itprop}) into the defining equation (\ref{propeq}) for $S(x,y)$,
we then obtain
\be
(- \Delta_{+} \Delta_{-} + m^2 ) \bar{G}(x,y) = \delta^4 (x,y),
\label{Geq}
\ee
with $\Delta_{\pm}$ given in eq. (\ref{deltapm}), and
\be
\Delta_{+} \Delta_{-} = ( i \partial_t + \Sigma_v)^2 +
\vec{e}\vec{e} - \frac{i}{L} \vec{\sigma} \vec{e} - \frac{9}{4L^2}.
\ee
(The derivatives act on $x$.) Note that $\vec{e}
\vec{e}$ is the Laplace-Beltrami operator on $S^3$:
\be
\vec{e}\vec{e} &=& {\rm g}^{- \frac12} \, \partial_{i} \,
g^{ \frac12} \, g^{i j} \, \partial_{j} \\ \nonumber &=&
\frac{\partial^2 }{\partial \mu^2 } + 2 \cot \mu
\frac{\partial }{\partial \mu } + \frac{1}{\sin^2 \mu}
\left( \frac{\partial^2 }{\partial \theta^2 } + \cot \theta
\frac{\partial }{\partial \theta } \right) + \frac{1}{\sin^2
\mu \sin^2 \theta } \frac{\partial^2 }{\partial \phi^2 } .
\label{l-b}
\ee
The $2 \times 2$ matrix structure of $\bar{G}(x,y)$ can be
expanded in the Pauli-matrix basis,
\be
\bar{G}(x,y) = G_1 (x,y) + \vec{\sigma} \vec{e}  \, G_2 (x,y),
\label{decomp}
\ee
where $G_1, G_2$ are scalar functions.

As in Minkowski space, the required invariance of the vacuum under
the symmetries of the metric leads to considerable simplifications
in the  space-time dependence of the propagator. Due to the symmetry
group of the spatial hypersphere, $SO(4)$, free propagation between
two points is equivalent to propagation of the same geodesic distance
from an arbitrarily fixed pole\footnote{This can be made explicit with
the help of the seven Killing vectors of $S^3 \times R$, which generate
the isometry group. They replace the ten Killing vectors of the
Poincar{\'e} group in flat Minkowski space.}. Taking the pole as the
origin of the polar coordinate system on $S^3$, the geodesic distance
simply becomes $\mu L$. Furthermore, the propagator will depend only
on time differences $t$ (the reference time is chosen to be zero), due
to the invariance of both metric and dynamics under time translations.

After inserting the decomposition (\ref{decomp}) of $G$ into eq.
(\ref{Geq}), we obtain two coupled equations for $G_{1,2}$:
\be
\left(\, -( i \partial_t + \Sigma_v)^2 - \vec{e}\vec{e} +
\frac{9}{4L^2} + m^2 \right) \,\,G_1(\mu,t) + \frac{i}{L} \vec{e}
\vec{e} \,\,G_2(\mu,t) = \delta^4 (x) \label{geq1}
\ee
and
\be
\left(\, -( i \partial_t + \Sigma_v)^2 - \vec{e}\vec{e} +
\frac{1}{4L^2} + m^2 \right) \,\, G_2(\mu,t) + \frac{i}{L} \,\,
G_1(\mu,t) = 0. \label{geq2}
\ee
(In the above derivation we used the Lie bracket relation $[e_a,e_b]
= \frac{2}{L} \epsilon_{ab}^{\,\,\,\,\,c} e_c$, see eqs. (\ref{gen1}),
(\ref{gen2}).) The remaining $\theta$ and $\phi$ dependence in eq.
(\ref{geq1}) is removed by integrating the first one over both angles
with their corresponding measure. (Eq. (\ref{geq2}) is already
independent of $\theta$ and $\phi$). The remaining explicit $\mu$
dependence can then be absorbed into the functions $\tilde{G}_{1,2}
(\mu,t) \equiv \sin \mu \,\,G_{1,2}(\mu,t)$ by writing the relevant
part of $\vec{e} \vec{e}$ as $L^{-2} \sin^{-1} \mu (1 +
\frac{d^2}{d\mu^2}) \sin \mu$ and the $\mu$-dependent part of the
delta function as $-\sin^{-1}\mu \delta ' (\mu)$ (a prime indicates
a derivative with respect to $\mu$). The Fourier transforms of
$\tilde{G}_{1,2}$ then satisfy simple algebraic equations,
\be
\left( (k_0 + \Sigma_v)^2 - k^2 - m^2 - \frac{5}{4 L^2} \right) \,\,
\tilde{G}_1 (k_0,k) &
+& \frac{i}{L} \left( k^2 - \frac{1}{L^2} \right) \,\,
\tilde{G}_2(k_0,k) =  \frac{ik}{2\pi L}  \nonumber \\
\left( (k_0 + \Sigma_v)^2 - k^2 - m^2 + \frac{3}{4 L^2} \right) \,\,
\tilde{G}_2 (k_0,k)
&-& \frac{i}{L} \,\, \tilde{G}_1(k_0,k) = 0,
\ee
which have the solutions (we introduce the abbreviations $k_{\pm}=
k \pm \frac{1}{2 L}, \,\, \delta = (k_0 + \Sigma_v)^2 - m^2$)
\be
\tilde{G}_2 (k_0,k) = \frac{-1}{4\pi L} \left( \frac{1}{\delta -
k_+^2} -
 \frac{1}{\delta - k_-^2} \right)
\ee
and
\be
\tilde{G}_1 (k_0,k) = - i L \, \left( \delta - k^2 +
\frac{3}{4 L^2} \right) \,
\tilde{G}_2 (k_0,k).
\ee
After transforming back to space-time and imposing Feynman boundary
conditions, we get
\be
\tilde{G}_1 (\mu,t) =  \frac{1}{4\pi L^2} \left( 2\cos \frac{\mu}{2}
I'(\mu,t) + \sin \frac{\mu}{2} I(\mu,t) \right) \,\, {\rm e}^{ i
\Sigma_v t}   \label{g1}
\ee
and
\be
\tilde{G}_2 (\mu,t) =  \frac{i}{2 \pi L} \sin \frac{\mu}{2} \,\,
{\rm e}^{ i \Sigma_v t} I(\mu,t) \label{g2}
\ee
in terms of the integral\footnote{$I(\mu,t)$ can be expressed in
terms of a Hankel function.}
\be
I(\mu,t) = \int \frac{dk_0}{2  \pi} \int \frac{d k}{2\pi} \,\,
\frac{{\rm e}^{i(k\mu L - k_0 t)} }{k_0^2 - k^2 - m^2 + i\epsilon},
\label{int}
\ee
which is the scalar Feynman propagator in two-dimensional Minkowski
space. The ultra-violet regularization of this integral and the
change in the pole structure of the integrand due to valence quarks
will be considered in appendix C. The full $(2 \times 2)$ propagator
can now be obtained by inserting eqs. (\ref{g1}) and (\ref{g2})
(divided by $\sin \mu$) into eq. (\ref{decomp}) and carrying out
the periodic sum:
\be
\bar{G}(x)  = \alpha (\sin \mu)^{-1} \, {\rm e}^{ i
(\frac{\mu}{2} \vec{ \sigma } \hat{r} + \Sigma_v t)}
\sum_{n=-\infty}^{+ \infty} (-1)^n [2 I_n' + \tan
\frac{\mu}{2} I_n], \label{hilf}
\ee
where $I_n(\mu,t) \equiv I (\mu + 2n \pi,t)$, the prime represents
a derivative with respect to $\mu$ and $\alpha \equiv 1/ 4 \pi L^2$.
It remains to evaluate the expression (\ref{itprop}) for the spinor
propagator, using the block anti-diagonal form of the Dirac operator
given in eq. (\ref{blockop}). As an intermediate step, we calculate
\be
\Delta_{\pm} \,\bar{G}(x) =  \alpha (\sin \mu)^{-1} \,
{\rm e}^{ i (\frac{\mu}{2} \vec{ \sigma } \hat{r} + \Sigma_v t)}
\sum_{n=-\infty}^{+ \infty} (-1)^n [2 \dot{I}_n' + \tan \frac{\mu}{2}
\dot{I}_n  \nonumber \\ \mp \frac{\vec{ \sigma } \hat{r}}{L} (2 I''_n
- \cot \frac{\mu}{2} I'_n)],
\ee
where the dot represents a time derivative. Inserted into
(\ref{itprop}), we obtain the quark propagator explicitly. It can be
brought into a familiar form by expanding the Dirac-matrix structure
in the $\gamma$-matrix basis:
\be
S(x) = {\rm e}^{ i ( \vec{ \Sigma } \hat{r} \frac{\mu}{2} +
\Sigma_v t)} [S_0 (\mu,t) \gamma_0 - S_1 (\mu,t) \hat{r} \vec{\gamma}
- S_2 (\mu,t)]. \label{fullS}
\ee
($\vec{\Sigma}$ is the Dirac spin matrix with the Pauli matrices on
the diagonal and should be distinguished from the self-energy.)
The three invariant scalar amplitudes $S_i$ are defined as
\be
S_0 (\mu,t) &=& -  i \alpha \, (\sin \mu)^{-1} \,
\sum_{n=-\infty}^{+ \infty} (-1)^n [2 \dot{I}_n' + \tan \frac{\mu}{2}
\dot{I}_n ], \label{S0} \\
S_1 (\mu,t) &=&  i \, \frac{\alpha }{L} \, (\sin \mu)^{-1} \,
\sum_{n=-\infty}^{+ \infty} (-1)^n [2 I''_n - \cot
\frac{\mu}{2} I'_n], \label{S1} \\
S_2(\mu,t) &=& m \alpha \, (\sin \mu)^{-1} \, \sum_{n=-\infty}^{+
\infty} (-1)^n [2 I_n' + \tan \frac{\mu}{2} I_n ] . \label{S2}
\ee
Eqs. (\ref{fullS}) -- (\ref{S2}) are the central result of this
appendix. The closed, analytical form of the fermion propagator on
$S^3 \times R$ may be useful also beyond the context of the present
paper.

The sums over all geodesic paths connecting the pole with $x$ ensure,
as mentioned above, the required spatial periodicity of the propagator.
This can now be easily checked explicitly. The $I_n(\mu,t)$ and their
derivatives, appearing in eqs. (\ref{S0}) -- (\ref{S2}), satisfy
\be
I_n (\mu + 2 \pi ,t) = I_{n+1} (\mu,t).
\ee
Inserted into the expression for the quark propagator, the expected
periodicity in $\mu$ follows immediately:
\be
S(\mu + 2 \pi, \theta, \phi, t) = (-1) \,\, {\rm e}^{ i \pi
\vec{ \Sigma } \hat{r} } \, S(\mu , \theta, \phi, t) = S(\mu ,
\theta, \phi, t) .
\ee

\section{Spectral representation and coincidence limits}

The sums over geodesic paths in the propagator functions (\ref{S0})
-- (\ref{S2}) can be rewritten alternatively as sums over the
spectrum of the Dirac operator. For many applications the spectral
representation, which we will now derive, is more transparent and
convenient. In numerical calculations it is particularly useful at
high densities, where the cell is small and paths which circle many
times around the sphere contribute significantly.

The $n$ dependence of the integrals $I_n$, defined after eq.
(\ref{hilf}),  and their derivatives resides exclusively in the
factors $(-1)^n \exp (2 i \pi n k L )$. Exchanging the order of
the sums over $n$ and the integrals over $k$ and using the
identity
\be
L \sum_{n = - \infty}^{\infty} (-1)^n {\rm e}^{2 i \pi n k L} =
\sum_{m = - \infty}^{\infty} \delta \left( k - \frac{2 m +1}{2 L}
\right) = \sum_{m = - \infty}^{\infty} \delta (k - k_m) ,
\label{sumid}
\ee
we can replace the sums over geodesic paths in the propagator by mode
sums over the discrete quark momenta,  eq. (\ref{mom}). After
performing the now trivial $k$ integrals this leads to the spectral
representation. Recall that the integrals over continuous $k$ in the
compact cell appeared since we took $\mu$  effectively in the interval
$\mu \in [ 0, \infty ]$, which allowed the use of continuous Fourier
methods in the derivation of the propagator. Consequently, we had to
add contributions from any number of (physically indistinguishable)
geodesic circles. The two equivalent ways of representing the
compactness of the unit cell are simply related by eq. (\ref{sumid}).

For the propagator functions $S_0$ and $S_2$ we now get the spectral
representations
\be
S_2 (\mu,t) &=& \frac{m \alpha}{ 2 \pi L} \, (\sin \mu)^{-1} \,
\sum_{n=-\infty}^{+ \infty} \int \frac{dk_0}{2  \pi}
\frac{{\rm e}^{i(k_n L \mu - k_0 t)} }{k_0^2 - k_n^2 - m^2 + i
\epsilon} \left[ 2  i k_n L  + \tan \frac{\mu}{2} \right]
\nonumber \\ &=& \frac{ i m}{4 V}  \sum_{n=1}^{ \infty} (\sin
\mu)^{-1} \left[ 2 k_n L \sin (k_n L \mu) - \cos (k_n L \mu) \tan
\frac{\mu}{2} \right] \frac{ {\rm e}^{-( i \omega_n + \epsilon)
|t| }}{\omega_n}   \label{S2spec}
\ee
and
\be
S_0 (\mu,t) &=& - \frac{  i}{m} \partial_t \, S_2 (\mu,t)
\nonumber \\ &=& - \frac{ i }{4 V}  \sum_{n=1}^{ \infty} (\sin
\mu)^{-1} \left[ 2 k_n L \sin (k_n L \mu) - \cos (k_n L \mu) \tan
\frac{\mu}{2} \right] \nonumber \\ & & \qquad \qquad \qquad \quad
\times \left[ \Theta (t) {\rm e}^{- i \omega_n t } - \Theta (-t)
{\rm e}^{ i \omega_n t }  \right].  \label{S0spec}
\ee

The evaluation of the constituent quark self-energies in section IV
requires the regularized coincidence limits of $S$. The time-ordering
ambiguity of the propagator in the coincidence limit is resolved as
usual by referring to the order in which the fields  appear in the
interaction lagrangian. Accordingly, we define
\be
S^{\pm} (0) = \lim_{x_0 \rightarrow 0 \pm} S(x_0, {\bf x} = 0) =
\tilde{S}_0^{\pm} \gamma_0 - \tilde{S}_2
\ee
and obtain
\be
\tilde{S}_0^{\pm} &=& \frac{  i}{4V}   \sum_{n= 1 }^{\infty}
(\mp 1 + f_n) D_n \Theta_{\epsilon} (\Lambda - k_n), \\
\tilde{S}_2 &=& \frac{  i m}{4V}  \sum_{n =1}^{\infty} \frac{D_n
(1 - f_n)}{\sqrt{k_n^2 + m^2}} \Theta_{\epsilon} (\Lambda - k_n) .
\ee
Here we have additionally accounted for the presence of the valence
quarks by adding the filling factor terms introduced in section IV.
We recognize again the appearance of the quark spectral properties,
which can alternatively be derived directly from the Dirac equation
(see appendix A).

\section{The free spin-0 field on $S^3 \times R$}

As a prerequisite for our discussion in sections V and VI, we will
derive here the mode decomposition of the free pion field (or any
real spin-0 field, in general) in the hyperspherical unit cell. We
denote the set of three spatial coordinates by ({\bf x}) and the
three generalized momentum quantum numbers (see below) by ({\bf k}),
and we suppress the trivial isospin indices.

Let us now expand the free field in the $S^3 \times R$ background as
\be
\phi (x) = \sum_{\bf k} \left( \eta_{\bf k} (x) a_{\bf k} + \eta_{\bf
k}^{*} (x) a_{\bf k}^{\dagger}  \right)
\ee
in terms of the complete set of solutions $\eta_{\bf k}$ of the free
Klein-Gordon equation\footnote{Note that we consider the minimally
coupled case. In general an additional term proportional to the Ricci
scalar of the metric can be added to the Klein-Gordon equation.} on
$S^3 \times R$, which generalize the usual plane waves of Minkowski
space:
\be
( \Box + m^2 ) \,\, \eta_{\bf k} (x) = \left( \frac{\partial^2}{\partial
t^2} - \vec{ e} \, \vec{ e} + m^2 \right) \eta_{\bf k} (x) = 0.
\label{kgeq}
\ee
($\vec{ e} \, \vec{ e}$ is the Laplace-Beltrami operator defined in
eq. (B8).) The $\eta$'s are chosen to be orthonormal in the
generalized scalar product
\be
(\eta_{\bf k},\eta_{\bf k'}) = - i \int_{S^3} d \mu ({\bf x}) \,\,
\eta_{\bf k} (x)\frac{ \stackrel{\leftrightarrow}{\partial} }{\partial
t} \eta_{\bf k'}^{*} (x) = \delta_{{\bf k},{\bf k'} }. \label{ornorm}
\ee
Note that the integral above does not include the time direction and
is restricted to the spatial hypersphere. One easily
shows that this definition of the scalar product is time independent.
In order to find the explicit form of the modes $\eta_{\bf k}$, we
first separate the time dependence,
\be
\eta_{\bf k} (x) = u_{\bf k} ({\bf x}) \frac{ {\rm e}^{- i \omega_{\bf
k} t }}{\sqrt{2 \omega_{\bf k} } },
\ee
and then solve the remaining static eigenvalue equation of the
Laplace-Beltrami operator:
\be
\vec{ e} \, \vec{ e} \,  u_{\bf k} ({\bf x})  = - \kappa_{\bf k}^2
u_{\bf k} ({\bf x}) \equiv (m^2 - \omega_{\bf k}^2 ) \,\, u_{\bf k}
({\bf x}), \label{eigeq}
\ee
where we defined $ \omega_{\bf k}^2 = \kappa_{\bf k}^2 + m^2$.
Separating further the dependence on the spatial coordinates, the
general solution of eq. (\ref{eigeq}) can be found to be
\be
u_{\bf k} ({\bf x}) = \frac{1}{\sqrt{L^3 h_n^{(l+1)} }} \sin^l \mu
\,\, C_n^{(l+1)} (\cos \mu) \,\, Y_{lm} (\theta, \phi).
\ee
Here the $C_n^{(\alpha)}$ are ultra-spherical Gegenbauer polynomials
\cite{Abram72}, the $Y_{lm}$ are spherical harmonics and the
$h_n^{(\alpha)}$ are normalization constants defined by
\be
h_n^{(\alpha)} = \frac{ \pi 2^{(1-2 \alpha)} \Gamma (n + 2
\alpha) }{n! (n + \alpha) \Gamma^2 ( \alpha) } \quad {\rm for}\quad
\alpha \neq 0.
\ee
The set ${\bf k}$ is now specified by the three quantum numbers
${\bf k} = \{ n,l,m \} $ within the range
\be
n &=& 0, 1, 2,... \nonumber \\
l &=& 0, 1, 2,...  \nonumber \\
m &=& -l, -l + 1, ..., l-1, l .
\ee
The eigenvalue corresponding to the eigenmode $u_{\bf k}$ is
\be
\kappa_{\bf k}^2 = \kappa_{n l}^2 = \frac{1}{L^2} (n+l) (n+l+2).
\ee
The static modes are orthogonal on $S^3$ and normalized to
\be
\int_{S^3} d \mu ({\bf x}) \,\, u_{\bf k} ({\bf x}) u_{\bf k'}^{*}
({\bf x}) = \delta_{{\bf k},{\bf k'} },
\ee
so that the time-dependent modes $\eta_{\bf k}$ satisfy the
orthonormality relation eq. (\ref{ornorm}). The standard equal-time
commutation relations
\be
[ \Phi ( {\bf x},t) , \Pi ( {\bf x}' ,t) ] =  i \delta^3 (
{\bf x},{\bf x}')
\ee
($\Pi = \partial_t \Phi$ is the canonical momentum conjugate to
$\Phi$, $\delta^3$ is the spatial part of the delta function defined
in appendix B) then insure the norm of the one-particle states:
\be
 [ a_{\bf k} , a_{\bf k'}^{\dagger}  ] &=& \delta_{{\bf k},{\bf k'} }
\nonumber \\
 \left[ a_{\bf k} , a_{\bf k'} \right] &=&  [ a_{\bf k}^{\dagger}  ,
a_{\bf k'}^{\dagger}  ] = 0,
\ee
so that
\be
<0|\phi^a(x)|\pi^b(k)> = \delta^{a b} \eta_{\bf k} (x). \label{pinorm}
\ee

\begin{figure}
\caption{ $-<\bar{q} q>^{1/3}$ as a function of density a) in the
hyperspherical unit cell (solid line) and for camparison b) in the
chemical potential approach of ref. \protect\cite{ber87} (dashed line).
Also shown is the constituent quark mass (dotted line).}
\label{fig1}
\end{figure}
\begin{figure}
\caption{ The free energy density (in units of $10^{10} MeV^4$) as
a function of the quark mass for four values of the baryon density.}
\label{fig2}
\end{figure}
 \begin{figure}
\caption{ $f_{\pi}$ as a function of density a) in the hyperspherical
unit cell (solid line) and for comparison b) in the chemical potential
approach of ref. \protect\cite{ber87} (dotted line). }
\label{fig3}
\end{figure}
\begin{figure}
\caption{ $-<\bar{q} q>^{1/3}$ and $f_{\pi}$ (in $MeV$) as a function
of density in the $B=0$ cell, which contains no valence quarks.
For comparison, the same curves in the $B=1$ cell from fig.s 1
and 3 are also shown (dotted line).}
\label{fig4}
\end{figure}

\end{document}